\documentclass[12pt]{article}
\usepackage{a4}
\usepackage{epsfig}

\def\brtoeq{\beta_{\rm e}}
\def\pt{p_{\rm T}}

\def\Z0{\mbox{Z}^0}
\def\LQ{\mbox{LQ}}
\def\X{\mbox{X}}
\def\ETMISS{E_{\rm T}\!\!\!\!\!\!\!/\;\;}
\def\ptmiss{\pt\!\!\!\!\!\!\!/\;\;}

\def\qqbar{\mbox{q}\overline{\mbox{q}}}

\newcommand{\inmath}[1] {\ifmmode#1\else$#1$\fi}
\newcommand{\definmath}[2] {\def#1{\ifmmode#2\else$#2$\fi}}
\newcommand{\yonetwo}{y_{12}}
\newcommand{\ytwothree}{y_{23}}
\newcommand{\mtlq}{m_{\rm T}^{\rm LQ}}

\definmath{\PWpm} {\mathrm{W}^{\pm}}      
\definmath{\Plp} {\ell^{+}}        
\definmath{\Plm} {\ell^{-}}        
\definmath{\Plpm}   {\ell^{\pm}}         
\definmath{\Pgtp} {\tau^{+}}        
\definmath{\Pgtm} {\tau^{-}}        
\definmath{\Pgtpm}   {\tau^{\pm}}         
\definmath{\Pgn}  {\nu}          
\definmath{\Pagn} {\overline{\nu}}     
\definmath{\Pq}      {\mathrm{q}}
\definmath{\Paq}  {\overline{\mathrm{q}}}
\definmath{\Pu}      {\mathrm{u}}
\definmath{\Pau}  {\overline{\mathrm{u}}}
\definmath{\Pd}      {\mathrm{d}}
\definmath{\Pad}  {\overline{\mathrm{d}}}
\definmath{\Ps}      {\mathrm{s}}
\definmath{\Pas}  {\overline{\mathrm{s}}}
\definmath{\Pc}      {\mathrm{c}}
\definmath{\Pac}  {\overlibstre{\mathrm{c}}}
\definmath{\Pb}      {\mathrm{b}}
\definmath{\Pab}  {\overline{\mathrm{b}}}
\definmath{\Pt}      {\mathrm{t}}
\definmath{\Pat}  {\overline{\mathrm{t}}}
\definmath{\Pap}  {\overline{\mathrm{p}}}
\definmath{\Pan}  {\overline{\mathrm{n}}}
\definmath{\PaD}  {\overline{\mathrm{D}}}
\definmath{\PaDz} {\overline{\mathrm{D}}^{0}}
\definmath{\PaB}  {\overline{\mathrm{B}}}
\definmath{\PaBz} {\overline{\mathrm{B}}^{0}}
\definmath{\PsDpm}   {\mathrm{D}^{\pm}_{\mathrm{s}}}  
\definmath{\PcgLpm}  {\Lambda^{\pm}_{\mathrm{c}}}  
\definmath{\PD} {\mathrm{D}}     
\definmath{\PDsm} {\mathrm{D_s^-}}     
\definmath{\PDst} {\mathrm{D}^{*}}     
\definmath{\PDstm} {\mathrm{D}^{*-}}     
\definmath{\PgLz} {\Lambda^{0}}        
\definmath{\Ppip} {\pi^+}
\definmath{\Ppim} {\pi^-}
\definmath{\Ppiz} {\pi^0}
\definmath{\PJPsi} {J/\Psi}
\definmath{\PKs} {{\rm K}_s}
\definmath{\PB} {{\rm B}}
\definmath{\PBs} {{\rm B_s}}
\definmath{\mumu} {\mu^+ \mu^-}
\definmath{\bb} {{\rm b\overline{b}}}

\newcommand{\epem}{\mbox{$\mathrm{e^+e^-}$}}

\newcommand {\ee}         {\mathrm{e}^+ \mathrm{e}^-}

\def\gappeq{\mathrel{ \rlap{\raise.5ex\hbox{$>$}}
                      {\lower.5ex\hbox{$\sim$}}  } }
\def\lappeq{\mathrel{ \rlap{\raise.5ex\hbox{$<$}}
                      {\lower.5ex\hbox{$\sim$}}  } }

\begin{document}
\begin{titlepage}
\begin{center}{\large   EUROPEAN ORGANIZATION FOR NUCLEAR RESEARCH
}\end{center}\bigskip
\begin{flushright}
       CERN-EP-2001-093   \\ 7 December 2001
\end{flushright}
\bigskip\bigskip\bigskip\bigskip\bigskip
\begin{center}{\huge\bf  
\boldmath
Search for Leptoquarks in
Electron-Photon Scattering at $\sqrt{s_{\rm ee}}$ up to $209$\,GeV at LEP
\unboldmath
}\end{center}\bigskip\bigskip
\begin{center}{\LARGE The OPAL Collaboration
}\end{center}

\bigskip

\bigskip\bigskip
\bigskip\begin{center}{\large  Abstract}\end{center}
Searches for first generation scalar and vector leptoquarks,
and for squarks in R-parity violating
SUSY models with the direct decay of the squark into Standard
Model particles, have been performed 
using $\ee$ collisions collected 
with the OPAL detector at LEP at e$^+$e$^-$ centre-of-mass energies
between 189 and 209\,GeV.
No excess of events is found over the expectation
from Standard Model background processes. Limits are computed
on the leptoquark
couplings for different values of the branching
ratio to electron-quark final states.  
\bigskip\bigskip\bigskip\bigskip
\bigskip\bigskip

\begin{center}{\large
(To be submitted to Physics Letters)
}\end{center}
\end{titlepage}

\begin{center}{\Large        The OPAL Collaboration
}\end{center}\bigskip
\begin{center}{
G.\thinspace Abbiendi$^{  2}$,
C.\thinspace Ainsley$^{  5}$,
P.F.\thinspace {\AA}kesson$^{  3}$,
G.\thinspace Alexander$^{ 22}$,
J.\thinspace Allison$^{ 16}$,
G.\thinspace Anagnostou$^{  1}$,
K.J.\thinspace Anderson$^{  9}$,
S.\thinspace Arcelli$^{ 17}$,
S.\thinspace Asai$^{ 23}$,
D.\thinspace Axen$^{ 27}$
% ...
}\end{center}\bigskip
\bigskip
$^{  1}$School of Physics and Astronomy, University of Birmingham,
Birmingham B15 2TT, UK
\newline
$^{  2}$Dipartimento di Fisica dell' Universit\`a di Bologna and INFN,
I-40126 Bologna, Italy
\newline
$^{  3}$Physikalisches Institut, Universit\"at Bonn,
D-53115 Bonn, Germany
\newline
% ...
$^{  5}$Cavendish Laboratory, Cambridge CB3 0HE, UK
\newline
% ...
$^{  9}$Enrico Fermi Institute and Department of Physics,
University of Chicago, Chicago IL 60637, USA
\newline
% ...
$^{ 16}$Department of Physics, Schuster Laboratory, The University,
Manchester M13 9PL, UK
\newline
$^{ 17}$Department of Physics, University of Maryland,
College Park, MD 20742, USA
\newline
% ...
$^{ 22}$Department of Physics and Astronomy, Tel Aviv University,
Tel Aviv 69978, Israel
\newline
$^{ 23}$International Centre for Elementary Particle Physics and
Department of Physics, University of Tokyo, Tokyo 113-0033, and
Kobe University, Kobe 657-8501, Japan
\newline
% ...
$^{ 27}$University of British Columbia, Department of Physics,
Vancouver BC V6T 1Z1, Canada
\newline
% ...

\section{Introduction}

The observed symmetry between the lepton and quark sectors in the Standard
Model is not yet understood, but could be interpreted as a hint for common 
underlying structures.  Consequently, many extensions of the Standard Model 
postulate the existence of leptoquarks (LQ), which are coloured spin 0 or 
spin~1 particles carrying both baryon (B) and lepton (L) quantum numbers.  
The Buchm\"uller-R\"uckl-Wyler (BRW) model~\cite{bib-buch}
adopted in this paper assumes lepton and baryon number conservation.
Two additional assumptions on the leptoquark couplings are made in the 
following: 
only the couplings $\lambda$ within one generation of fermions
are assumed to be non-zero (to respect lepton flavour conservation), and 
only 
the case of chiral couplings is considered, i.e.~it is assumed that the product
$\lambda_{\rm R}\lambda_{\rm L}$ of couplings to left-handed and right-handed
leptons vanishes (to respect lepton universality).
With the latter assumption the branching ratio $\brtoeq$ of leptoquarks to final states
with a charged lepton is restricted, as shown in Table~\ref{tab-lqstates}.

\begin{figure}[htb]
\vspace{6.5cm}
   \begin{center}
          \epsfxsize=10cm
          \epsffile{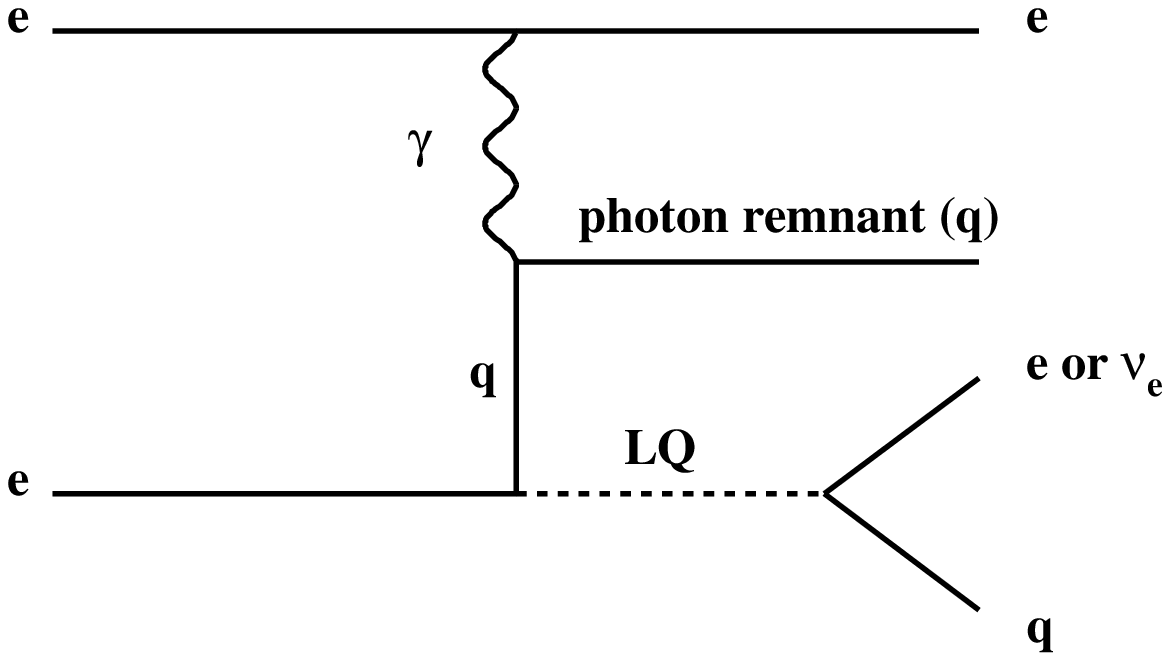}
\caption{Schematic diagram of the $s$-channel
production of a leptoquark in electron-photon
scattering. The photon is radiated by one of the LEP beams, fluctuates into 
a hadronic object, and one of the quarks interacts with an electron from 
the other beam.}
\label{dominantdiagram.fig}
\vspace{20.5cm}
\end{center}
\end{figure}

This paper presents the results of a search for single production
of leptoquarks in $\epem$ collisions with the OPAL detector at LEP.
Leptoquarks would be produced by an electron-quark\footnote{Charge 
conjugation is implied throughout this paper for all particles, 
e.g.~positrons are also referred to as electrons.} fusion.
In the dominant diagram, which is shown schematically
in Figure~\ref{dominantdiagram.fig},
the quark is the result of a 
process where one of the incoming leptons radiates
a photon which subsequently fluctuates into a hadronic 
state~\cite{bib-theo}.
Under the above assumptions on the leptoquark couplings,
only first generation leptoquarks coupled to electrons
could be produced in electron-photon scattering, and they
would decay into either an electron and a quark or into a 
neutrino and a quark. 
Therefore the main signature of single leptoquark events is
one hadronic jet balanced
in the transverse plane either by one isolated electron
or by missing transverse energy due to the neutrino.
The hadronic photon remnant would disappear down the beam-pipe 
or add some activity in the forward region of the detector.

\renewcommand{\arraystretch}{1.2}
\begin{table}[htbp]
  \begin{center}
    \begin{tabular}{|c|c|c|cc|c|}\hline   
                                &        &     & \multicolumn{2}{|c|}{coupling and} & \\ 
  scalar LQ ($\tilde{\mbox{q}}$) & charge & $F$ & \multicolumn{2}{|c|}{decay mode}   & $\brtoeq$ \\ 
\hline
\hline
  S$_{\rm 0}$ (or ${\rm \tilde{d}}_R$) & $-$1/3 & 2 & 
  $\begin{array}{@{}c@{}} \lambda_{\rm L}: \\ \lambda_{\rm R}: \end{array}$ &
  \begin{tabular}{@{}c@{}} e$^-_{\rm L}$u, $\nu_{\rm L}$d \\ e$^-_{\rm R}$u \end{tabular} &
  \begin{tabular}{@{}c@{}} 1/2 \\ 1 \end{tabular} \\
\hline
  \~S$_{\rm 0}$ & $-$4/3 & 2 & 
  $\lambda_{\rm R}$: & 
  e$^-_{\rm R}$d & 
  1 \\ 
\hline
  \~S$_{\rm 1/2}$ (or ${\rm \bar{\rm \tilde{d}}_L}$) & +1/3 & 0 &  
  $\lambda_{\rm L}$: & 
  $\nu_{\rm L}\bar{\rm d}$ & 
  0 \\
  \~S$_{\rm 1/2}$ (or ${\rm \bar{\rm \tilde{u}}_L}$) & $-$2/3 & 0 &  
  $\lambda_{\rm L}$: & 
  e$^-_{\rm L}\bar{\rm d}$ & 
  1 \\ 
\hline
  S$_1$ & 
  \begin{tabular}{@{}c@{}} +2/3 \\ $-$1/3 \\ $-$4/3 \end{tabular} &
  2 &
  $\begin{array}{@{}c@{}} \lambda_{\rm L}: \\ \lambda_{\rm L}: \\ \lambda_{\rm L}: \end{array}$ &
  \begin{tabular}{@{}c@{}} $\nu_{\rm L}$u \\
                           $\nu_{\rm L}$d, e$^-_{\rm L}$u \\
                           e$^-_{\rm L}$d \end{tabular} &
  \begin{tabular}{@{}c@{}} 0 \\ 1/2 \\ 1 \end{tabular} \\
\hline
  S$_{1/2}$ &
  \begin{tabular}{@{}c@{}} $-$2/3 \\ \\ $-$5/3 \end{tabular} &
  0 &
  $\begin{array}{@{}c@{}} \lambda_{\rm L}: \\ \lambda_{\rm R}: \\ 
                          \lambda_{\rm L}: \\ \lambda_{\rm R}: \end{array}$ &
  \begin{tabular}{@{}c@{}} $\nu_{\rm L}\bar{\rm u}$ \\
                           e$^-_{\rm R}\bar{\rm d}$ \\
                           e$^-_{\rm L}\bar{\rm u}$ \\
                           e$^-_{\rm R}\bar{\rm u}$ \end{tabular} &
  \begin{tabular}{@{}c@{}} 0 \\ 1 \\ 1 \\ 1 \end{tabular} \\
\hline
\multicolumn{6}{c}{}\\
\hline
            &        &     & \multicolumn{2}{|c|}{coupling and} & \\ 
 vector LQ  & charge & $F$ & \multicolumn{2}{|c|}{decay mode}   & $\brtoeq$ \\ 
\hline\hline
  V$_{1/2}$ &
  \begin{tabular}{@{}c@{}} $-$1/3 \\ \\ $-$4/3 \end{tabular} &
  2 &
  $\begin{array}{@{}c@{}} \lambda_{\rm L}: \\ \lambda_{\rm R}: \\ 
                          \lambda_{\rm L}: \\ \lambda_{\rm R}: \end{array}$ &
  \begin{tabular}{@{}c@{}} $\nu_{\rm L}$d \\
                           e$^-_{\rm R}$u \\
                           e$^-_{\rm L}$d \\
                           e$^-_{\rm R}$d \end{tabular} &
  \begin{tabular}{@{}c@{}} 0 \\ 1 \\ 1 \\ 1 \end{tabular} \\
\hline
  \~V$_{1/2}$ & 
  \begin{tabular}{@{}c@{}} +2/3 \\ $-$1/3 \end{tabular} &
  2 &
  $\begin{array}{@{}c@{}} \lambda_{\rm L}: \\ \lambda_{\rm L}: \end{array}$ &
  \begin{tabular}{@{}c@{}} $\nu_{\rm L}$u \\
                           e$^-_{\rm L}$u \end{tabular} &
  \begin{tabular}{@{}c@{}} 0 \\ 1 \end{tabular} \\
\hline
  V$_0$ & $-$2/3 & 0 &
  $\begin{array}{@{}c@{}} \lambda_{\rm L}: \\ \lambda_{\rm R}: \end{array}$ &
  \begin{tabular}{@{}c@{}} e$^-_{\rm L}\bar{\rm d}$, $\nu_{\rm L}\bar{\rm u}$ \\
                           e$^-_{\rm R}\bar{\rm d}$ \end{tabular} &
  \begin{tabular}{@{}c@{}} 1/2 \\ 1 \end{tabular} \\
\hline
  V$_1$ &
  \begin{tabular}{@{}c@{}} +1/3 \\ $-$2/3 \\ $-$5/3 \end{tabular} &
  0 &
  $\begin{array}{@{}c@{}} \lambda_{\rm L}: \\ \lambda_{\rm L}: \\ \lambda_{\rm L}: \end{array}$ &
  \begin{tabular}{@{}c@{}} $\nu_{\rm L}\bar{\rm d}$ \\
                           e$^-_{\rm L}\bar{\rm d}$, $\nu_{\rm L}\bar{\rm u}$ \\
                           e$^-_{\rm L}\bar{\rm u}$ \end{tabular} &
  \begin{tabular}{@{}c@{}} 0 \\ 1/2 \\ 1 \end{tabular} \\
\hline
  \~V$_0$ & $-$5/3 & 0 & 
  $\lambda_{\rm R}$: & 
  e$^-_{\rm R}\bar{\rm u}$ & 
  1 \\ 
\hline
  \end{tabular}
    \caption{The first generation scalar (S) leptoquarks/squarks and vector (V) 
leptoquarks in the BRW model according to the nomenclature in~\cite{bib-nom} 
with their electric charge in units of $e$ and fermion
number $F=L+3B$.  For each possible non-zero coupling $\lambda$ the
decay modes and the corresponding branching ratio $\brtoeq$ 
for the decay into an electron and a quark are also listed.  The restrictions on the
values of $\brtoeq$ arise from the assumption of chiral couplings.}
    \label{tab-lqstates}
  \end{center}
\end{table}
\renewcommand{\arraystretch}{1.0}

In supersymmetric models with R-parity violation, scalar quarks (squarks) have the same 
production mechanism as some leptoquarks.  The topology of events with
single squark production depends on the squark decays; if R-parity conserving
decays can be neglected, the results on the left-handed couplings for 
leptoquarks presented in this paper apply
also to the corresponding squark states.  In the following, leptoquarks and the 
corresponding squark states with R-parity violating decays 
are generically referred to as leptoquarks.

Experiments at LEP~\cite{bib-karina,bib-leplq},
HERA~\cite{bib-heralq}, and the 
Tevatron~\cite{bib-tevatronlq} have
searched for leptoquarks. 
We present a search for leptoquarks with $M_{\rm LQ}>80$\,GeV
in electron-photon scattering using
data corresponding to an integrated luminosity of 
$612.3$\,pb$^{-1}$ 
at $\epem$ centre-of-mass energies between 189 and 209\,GeV. 
The data sample is split into eight bins with average $\epem$ centre-of-mass energies
of 
$188.6\,{\rm GeV}$,
$191.6\,{\rm GeV}$,
$195.5\,{\rm GeV}$,
$199.5\,{\rm GeV}$,
$201.6\,{\rm GeV}$,
$204.9\,{\rm GeV}$,
$206.5\,{\rm GeV}$, and
$208.0\,{\rm GeV}$.
This paper extends the previous OPAL analysis~\cite{bib-karina}
to the highest LEP centre-of-mass energies and uses
refined experimental techniques.  The data used in~\cite{bib-karina} 
are included in this analysis and thus the 
results from~\cite{bib-karina} are superseded.

\section{The OPAL Detector}
\label{sec-opal}  
The OPAL detector is described in detail elsewhere~\cite{opaltechnicalpaper}. 
It is a multipurpose apparatus
with almost complete solid angle coverage.
The central detector consists of two layers of
silicon micro-strip detectors~\cite{simvtx} and
a system of gas-filled tracking chambers in a 0.435\,T solenoidal magnetic field
which is parallel to the beam axis.

A lead-glass electromagnetic calorimeter 
with a presampler is located outside the magnet coil. 
In combination with the forward calorimeters,
the forward scintillating tile counter~\cite{bib-llpaper},
and the silicon-tungsten luminometer~\cite{bib-siw}, 
a geometrical acceptance is provided down 
to 25\,mrad from the beam direction.
The silicon-tungsten luminometer
measures the integrated luminosity using small-angle Bhabha
scattering events~\cite{lumino}.
The magnet return yoke is instrumented for hadron calorimetry, and
is surrounded by several layers of muon chambers.

\section{Monte Carlo Simulation}
\label{sec-mc}
The Monte Carlo simulation of the process $\mbox{e}^+\mbox{e}^-\to \LQ +\X+{\rm e}$
and the calculation of the production cross-section, defined as the sum
of the cross-sections for the particle and antiparticle states, are
performed with the program {\tt ERATO-LQ}~\cite{bib-erato}.
The hadronisation of the leptoquark decay products and the 
hadronic photon remnant is performed using {\tt JETSET}~\cite{bib-pythia}.
Details of the Monte Carlo simulation of single leptoquark
production and the calculation of the total cross-section
can be found in~\cite{bib-karina}.
Samples of 3000 signal events are generated for both scalar and 
vector leptoquark states, for both eq and $\nu$q final states, 
for $\epem$ centre-of-mass energies $\sqrt{s_{\rm ee}}$ of $200$ and $208\,{\rm GeV}$,
and for scaled leptoquark masses $x_{\rm LQ} = m_{\rm LQ}/\sqrt{s_{\rm ee}}$
of $x_{\rm LQ} = 0.42$, $0.53$, $0.63$, $0.74$, $0.85$, $0.90$, $0.95$, and $0.98$,
making a total of 64 samples.
For a given scaled leptoquark mass
the event properties depend
only weakly on the centre-of-mass energy.

To study Standard Model background processes, 
two-fermion hadronic events ($\ee\rightarrow \qqbar(\gamma)$) are 
simulated with {\tt PYTHIA 5.722}~\cite{bib-pythia}, while the $\ee\rightarrow\tau^+\tau^-$
and $\ee\rightarrow\ee$ processes are generated with 
{\tt KORALZ 4.02}~\cite{bib-koralz} and
{\tt BHWIDE}~\cite{bib-bhwide}, respectively.
Deep inelastic hadronic two-photon background events in the
range $Q^2>4.5$~GeV$^2$, including charged current 
deep inelastic scattering (CC DIS) events, are simulated with 
{\tt HERWIG 5.8}~\cite{bib-herwig}, while
{\tt PHOJET 1.10}~\cite{bib-phojet} is used to generate hadronic two-photon events
in the range $Q^2<4.5$~GeV$^2$.
For leptonic two-photon events, {\tt Vermaseren}~\cite{bib-vermaseren} is used.
Final states with four-fermion
production are simulated with {\tt grc4f}~\cite{bib-grc4f}.

All Monte Carlo events are passed through a full detector 
simulation~\cite{bib-GOPAL} and the same reconstruction algorithms as the real data.

\section{Event Analysis}
\label{sec-evsel}
We search for events with one hadronic jet and either an 
electron or missing energy balancing the transverse momentum of this jet.
The analysis uses tracks reconstructed in the central tracking devices and
clusters measured in the calorimeters and forward detectors.
The selection of tracks and clusters is similar to that
used in previous OPAL analyses.
In addition to quality requirements~\cite{bib-karina},
tracks must have more than 20 hits in the central jet chamber,
more than half the number of hits expected 
given the polar angle of the track and the geometry
of the jet chamber, and a
transverse momentum with respect to
the beam direction of more than 120\,MeV.
Calorimeter clusters must pass energy threshold cuts to
suppress noise.
To avoid double counting of particle energies, a matching algorithm
between tracks and clusters is applied~\cite{bib-mt}.

Jets are reconstructed from the tracks and clusters using
the Durham~\cite{bib-durham} and cone~\cite{bib-conejet} algorithms.
The $y$ parameter value $\yonetwo$ ($\ytwothree$) where the transition from a 
one- to two-jet (two- to three-jet) event occurs is determined with the 
Durham algorithm.
These parameters provide information on the event shape in the following analysis.
A cone jet finder~\cite{bib-conejet},
requiring a minimum energy of 15\,GeV in a cone of half-angle 1.0 radian
is used
to provide information on jet directions and energies.
No requirement on the number of jets is made.  

\subsection{The Electron Plus Hadronic Jet Channel}
Tracks are identified as electrons if more than
20 measurement samples can be used to determine the specific ionisation energy loss,
${\mathrm d}E/{\mathrm d}x$, if the ${\mathrm d}E/{\mathrm d}x$ probability~\cite{bib-dedxprobability} 
for the electron hypothesis exceeds $1\%$, and if 
the ratio of the electron energy measured
in the electromagnetic calorimeter to the track momentum 
is between $0.7$ and $2.0$. 
In each event, the electron with the largest momentum is 
assumed to be the one from leptoquark decay. 

Candidate leptoquark events are then selected based on
the following cuts:
\begin{itemize}
\item[{EQ1)}]
The event must contain more than five tracks, and the energy
measured in the hadron calorimeter must exceed 1\,GeV.  
\item[{EQ2)}]
The ratio $\ETMISS/E_{\rm vis}$ of the missing transverse energy $\ETMISS$ and the visible
energy $E_{\rm vis}$ has to be smaller than $0.2$.
This cut mainly reduces background from 4-fermion and two-photon events.
\item[{EQ3)}]
The event must contain an identified electron, and its energy $E_{\rm e}$ 
has to be larger than $0.2\sqrt{s_{\rm ee}}$, where $\sqrt{s_{\rm ee}}$
denotes the $\epem$ centre-of-mass energy.
The energy of the cone jet containing the most energetic
electron must not exceed the electron energy by more than $12\%$ of the 
centre-of-mass energy.
This ensures that isolated, energetic electrons are selected.
\item[{EQ4)}]
At least one cone jet not containing the electron is required, and the leptoquark
is reconstructed from the electron and the most energetic other cone jet (quark jet).
The quantity $\cos\theta^*_{\rm e}$ is the cosine of the 
helicity angle between the electron in the leptoquark rest frame and the leptoquark
flight direction in the laboratory frame.
Since high-mass leptoquarks will be almost at rest in the 
laboratory, the finite energy and momentum resolution of the 
detector will produce an apparent leptoquark direction close 
to the jet or the electron direction. Therefore the 
distribution of $\cos\theta^*_{\rm e}$ will be strongly peaked at $\pm 1$.
To remove small-angle Bhabba events in which one electron
produces a shower in the tracking chambers, the multiplicity
of the quark jet (defined as the number of tracks
plus unassociated clusters) is required to be larger than 10
for $|\cos\theta^*_{\rm e}|>0.95$ and $|\cos\theta_{\rm e}|>0.8$, where
$\theta_{\rm e}$ denotes the electron polar angle with respect to the outgoing
electron beam\footnote{The OPAL coordinate 
system is defined as a right-handed Cartesian coordinate system, with the 
$x$ axis pointing in the plane of the LEP collider towards the 
centre of the ring, and the $z$ axis in the direction of the 
outgoing beam electrons.}.
\end{itemize}
The above cuts are applied for all centre-of-mass energies and leptoquark 
masses and states.
To reject background further, an 
$x_{\rm LQ}$-dependent likelihood is constructed, where the leptoquark mass is reconstructed as the 
invariant mass of the electron and the quark jet.  The likelihood uses as inputs:
\begin{itemize}
\item
the charge-signed cosine of the electron polar angle, $-q_{\rm e}\cos\theta_{\rm e}$,
\item
the quantity $\cos\theta^*_{\rm e}$ defined above,
\item
the logarithm of $\ytwothree$ from the Durham jet finder,
\item
the cosine of the polar angle of a third jet in the event,
if one exists\footnote{Events in which no third cone jet is found are 
collected in a separate bin.}, multiplied by the sign of $\cos\theta_{\rm e}$, and
\item
the ratio $\ETMISS/E_{\rm vis}$.
\end{itemize}
Monte Carlo samples that are statistically independent of those used in the efficiency
determination are used to obtain the probability density to construct the likelihood function.
Two-fermion events, four-fermion events, and two-photon events are considered as 
separate background hypotheses.  
       The signal reference distributions for the likelihood
       for a given event should ideally be derived from
       Monte Carlo simulation of a leptoquark with a mass
       corresponding to the measured electron-jet mass
       of that event.  Since the scaled electron-jet mass 
       $x_{\rm LQ}$ of the event does not in general correspond to one of the
       signal 
       Monte Carlo masses, bin-by-bin interpolation 
       from the two Monte Carlo samples with $x_{\rm LQ}$ bracketing that
       of the measured event is used to determine the signal
       reference histograms used for its likelihood.
For each event, 
the signal likelihood is calculated from the average
of the scalar and vector leptoquark distributions at the reconstructed leptoquark mass.
Finally, events are accepted if they pass a cut on the likelihood value ${\cal L}>0.8$.

In Table~\ref{cutfloweq.table}, the number
of data events and the expected number of Standard Model background events 
are shown after each selection cut.
The selection efficiencies for two different scalar 
and vector leptoquark masses at 200\,GeV $\epem$ centre-of-mass energy are also given. 
       The likelihood interpolation technique 
       described above was checked at each 
       leptoquark mass simulated in the Monte Carlo
       by comparing the efficiency obtained when 
       using reference histograms derived directly
       from the Monte Carlo at that mass to that
       obtained when the histograms were derived
       by interpolating the next two nearest simulated
       mass points.  It was found that the interpolation
       method degrades the efficiency by about 15\%
       for low-mass leptoquarks, while it has a very small
       effect for high-mass leptoquark.  The efficiency
       for leptoquark detection is therefore reduced to account
       for this effect.

\renewcommand{\arraystretch}{1.0}
\begin{table}[htbp]
  \begin{center}
    \begin{tabular}{|c||c|c|c|c|c|}
      \hline
               & \multicolumn{5}{|c|}{selection steps (cumulative cuts)} \\
        events & (EQ1) & (EQ2) & (EQ3) & (EQ4) & ${\cal L}>0.8$ \\
      \hline
       \multicolumn{6}{c}{signal efficiencies:} \\
       \hline
 scalar LQ, $m_{\rm LQ}=106.0\,{\rm GeV}$
 & $72.6\%$ 
 & $55.0\%$
 & $39.8\%$
 & $39.7\%$
 & $(15.6 \pm 1.1)\%$  
 \\
       \hline
 scalar LQ, $m_{\rm LQ}=190.5\,{\rm GeV}$
 & $79.4\%$ 
 & $69.3\%$
 & $57.3\%$
 & $56.7\%$
 & $(49.2 \pm 2.7)\%$      
 \\
       \hline
 vector LQ, $m_{\rm LQ}=106.0\,{\rm GeV}$
 & $66.5\%$ 
 & $46.8\%$
 & $32.4\%$
 & $32.4\%$
 & $\enspace(9.8 \pm 0.8)\%$      
 \\
       \hline
 vector LQ, $m_{\rm LQ}=190.5\,{\rm GeV}$
 & $70.6\%$ 
 & $61.9\%$
 & $51.5\%$
 & $51.1\%$
 & $(47.7 \pm 2.7)\%$      
 \\
       \hline
       \multicolumn{6}{c}{expected background events:} \\
       \hline
 $\epem\to{\rm 2\, fermions}$
 & $\enspace46391\phantom{.0}$ 
 & $17682\phantom{.0}$ 
 & $\enspace94.0$ 
 & $\enspace91.5$ 
 & $\enspace9.1\pm\enspace2.9$  
 \\
       \hline
 $\epem\to{\rm 4\, fermions}$
 & $\enspace12800\phantom{.0}$ 
 & $\enspace5111\phantom{.0}$ 
 & $431\phantom{.0}$ 
 & $428\phantom{.0}$ 
 & $23.7\pm\enspace7.4$  
 \\
       \hline
 two-photon
 & $765676\phantom{.0}$ 
 & $\enspace5702\phantom{.0}$ 
 & $\enspace71.0$ 
 & $\enspace67.3$ 
 & $11.9\pm\enspace3.9$  
 \\
       \hline
 charged-current DIS
 & $\enspace\enspace\enspace\enspace25.8$ 
 & $\enspace\enspace\enspace\enspace0.1$ 
 & $<0.1$ 
 & $<0.1$ 
 & $<0.1$ 
 \\
       \hline
       \hline
 total expected background
 & $824892\phantom{.0}$ 
 & $28496\phantom{.0}$ 
 & $596\phantom{.0}$ 
 & $587\phantom{.0}$ 
 & $44.7\pm14.0$  
 \\
       \hline
       \multicolumn{6}{c}{observed events:} \\
       \hline
 data
 & $735325\phantom{.0}$
 & $28888\phantom{.0}$
 & $562\phantom{.0}$
 & $536\phantom{.0}$
 & $43$
 \\
       \hline
    \end{tabular}
    \caption{Selection efficiencies in the eq~channel.
The listed efficiencies are for $\sqrt{s_{\rm ee}}=200\,{\rm GeV}$.
The remaining number of data
events and the expected number of background events are also
listed after each selection cut.
All event numbers are quoted after a loose event preselection in which 
more than 2 tracks, more than $0.1\,{\rm GeV}$ total energy in the 
hadron calorimeter, and an identified electron are required.
The errors are the total systematic uncertainties including the Monte
Carlo statistical error.  The component which 
is not due to Monte Carlo statistics is correlated between the separate 
contributions to the background.  The determination of the systematic errors 
is discussed in Section~\ref{systunc.sec}.}
    \label{cutfloweq.table}
  \end{center}
\end{table}

\begin{figure}[htb]
  \begin{center}
    \unitlength 0.5814pt
    \begin{picture}(500,600)
      \put(-130,390){\epsfig
{file=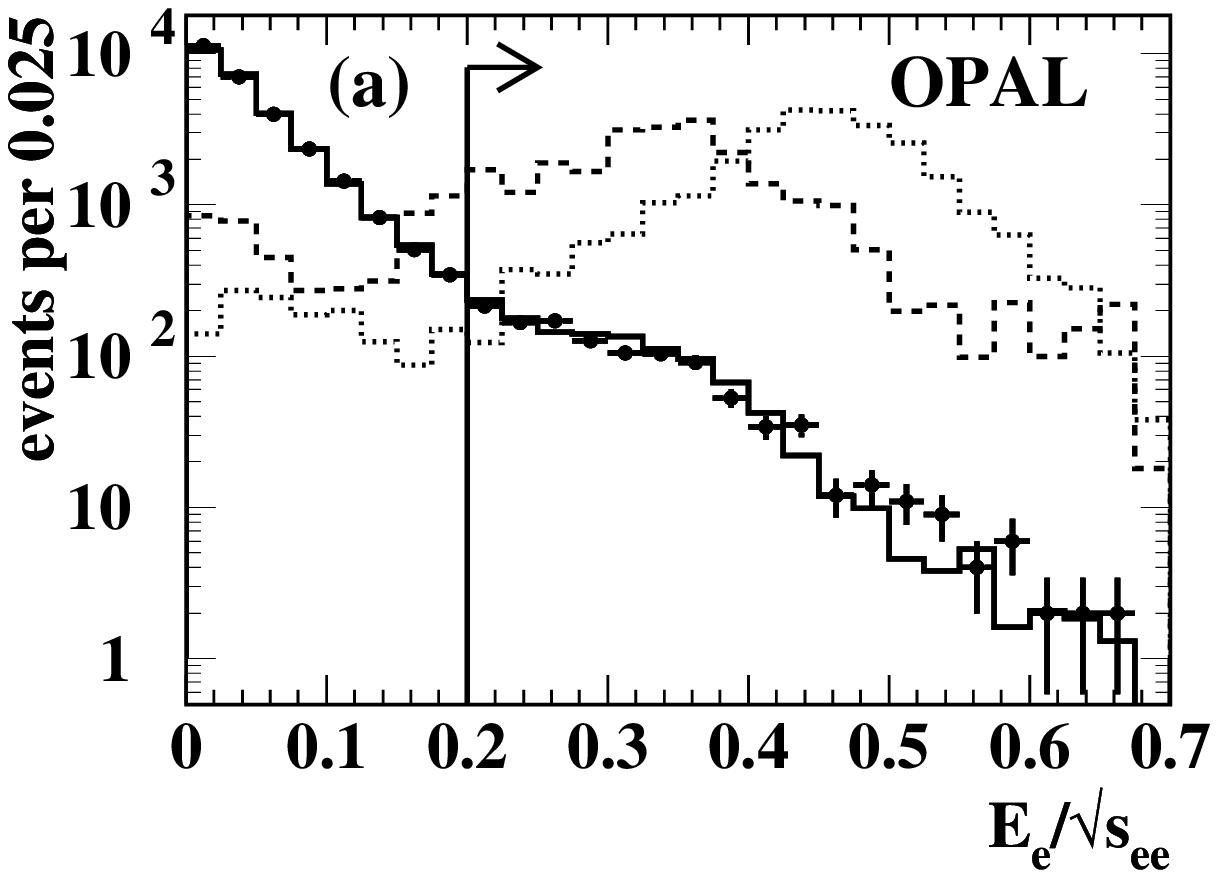,width=400pt,bbllx=20pt,bblly=305pt,bburx=640pt,bbury=572pt}}
      \put(270,390){\epsfig
{file=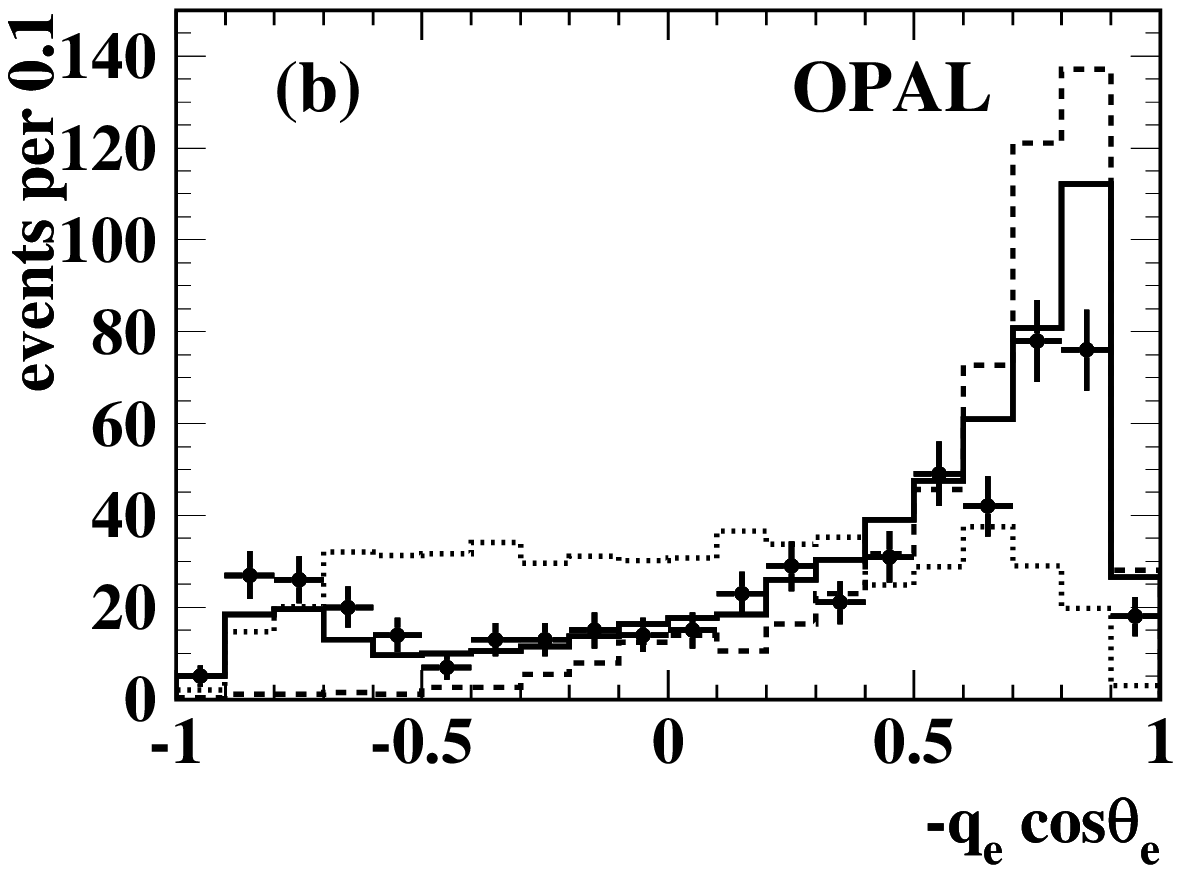,width=400pt,bbllx=20pt,bblly=305pt,bburx=640pt,bbury=572pt}}
      \put(-130,100){\epsfig
{file=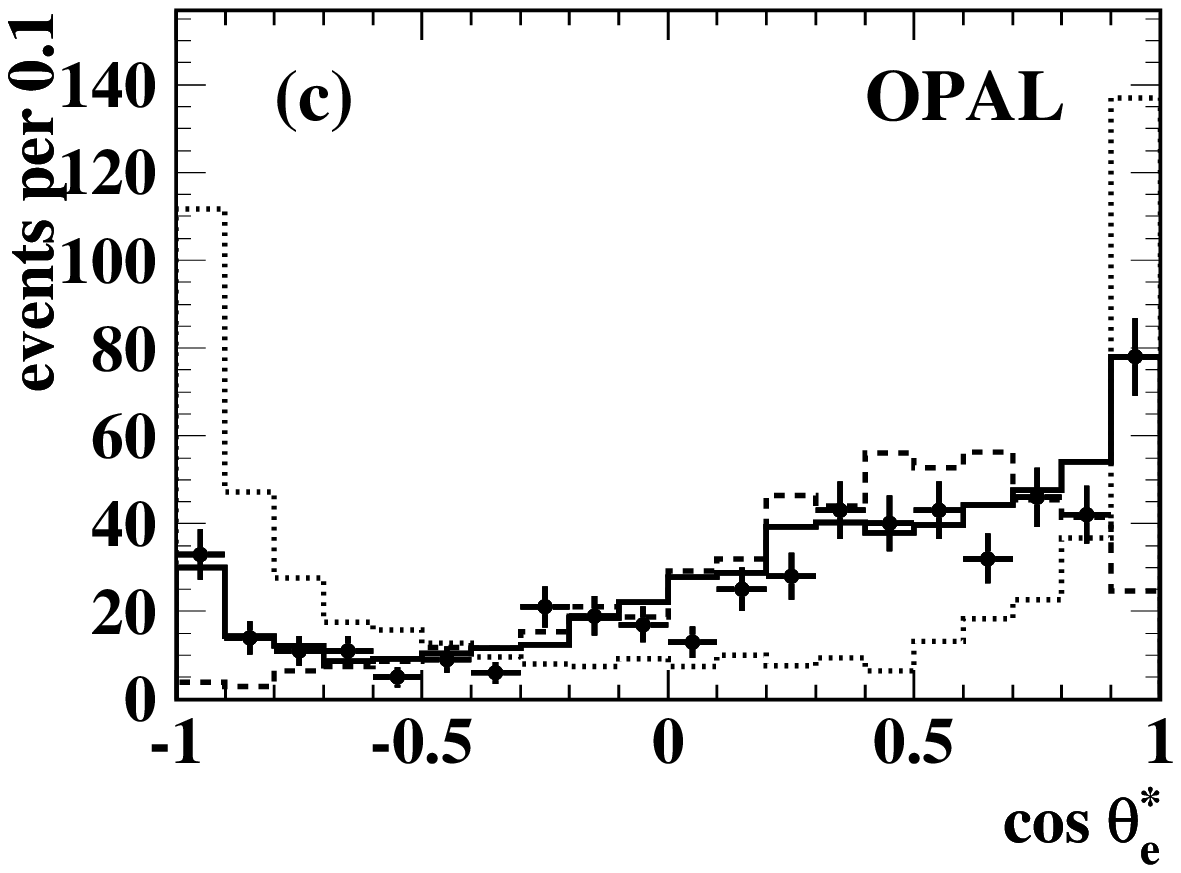,width=400pt,bbllx=20pt,bblly=305pt,bburx=640pt,bbury=572pt}}
      \put(270,100){\epsfig
{file=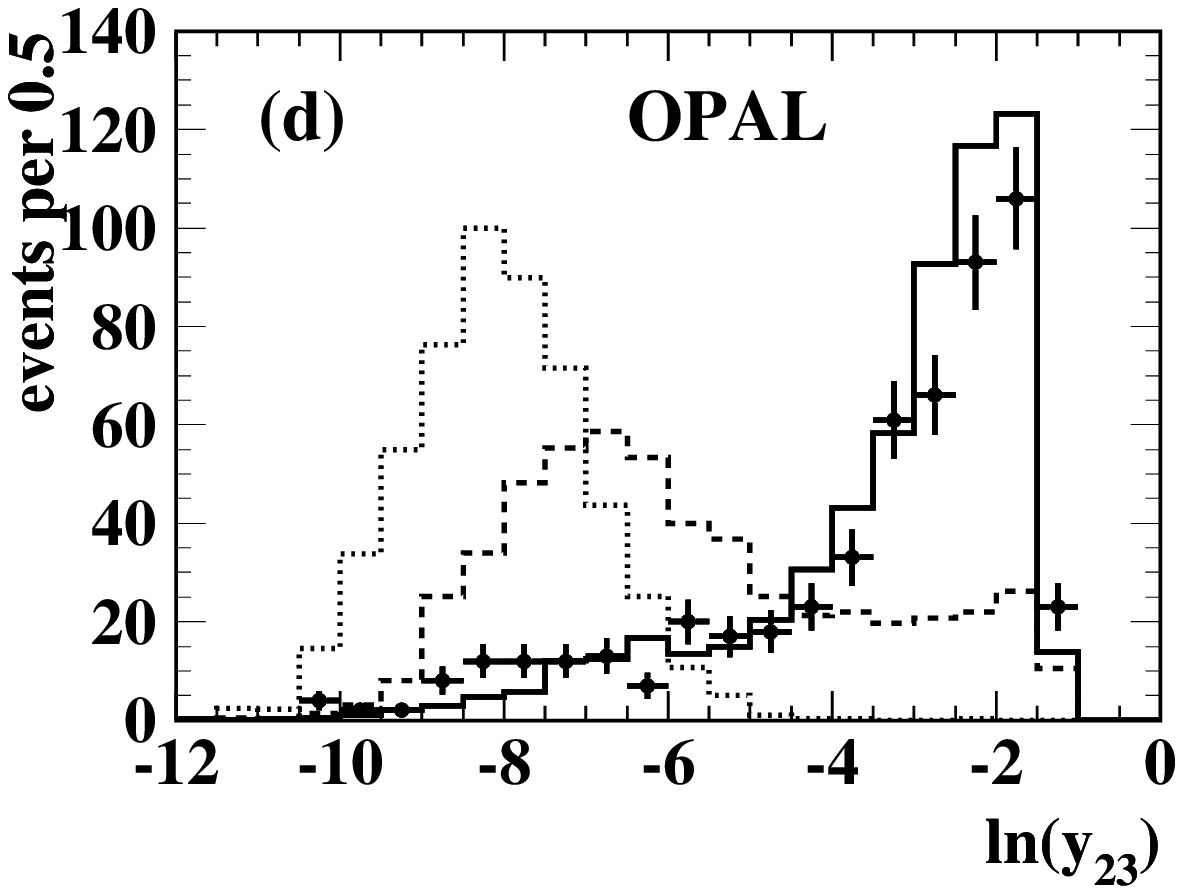,width=400pt,bbllx=20pt,bblly=305pt,bburx=640pt,bbury=572pt}}
      \put(-130,-190){\epsfig
{file=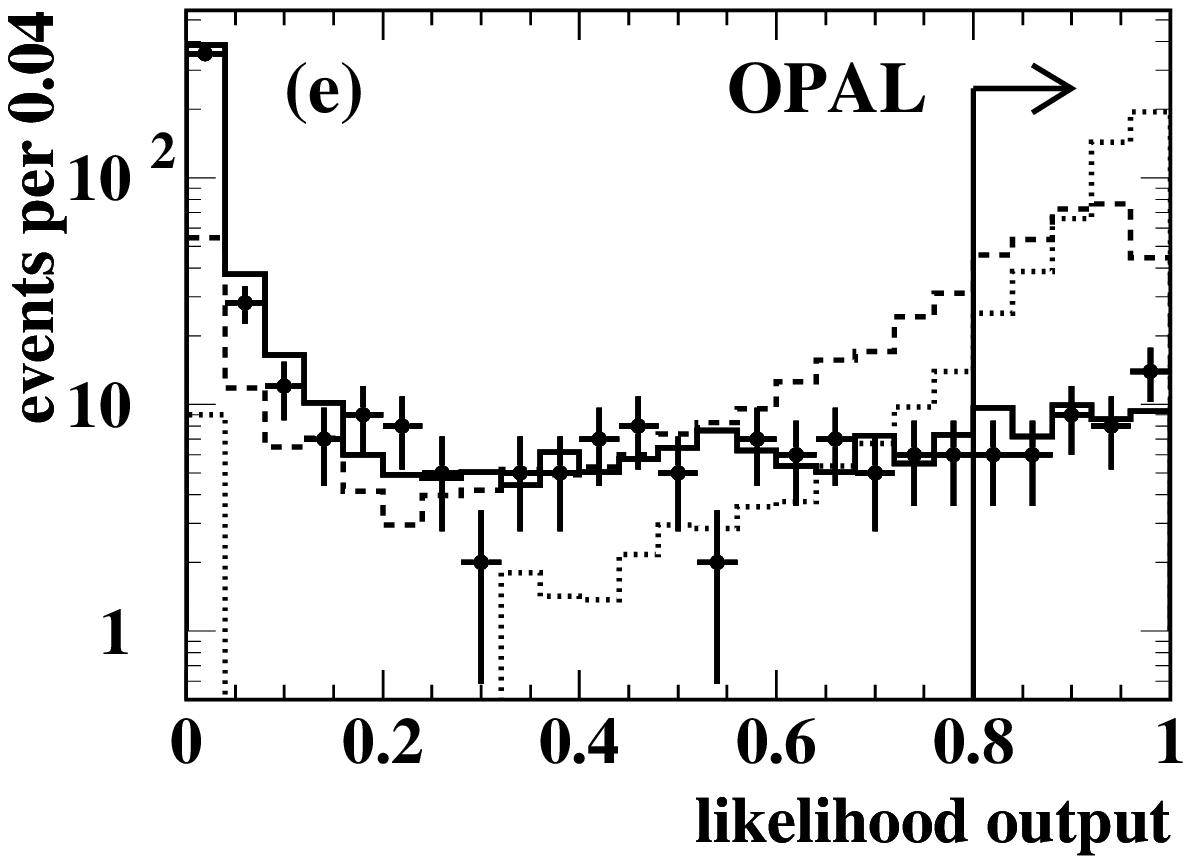,width=400pt,bbllx=20pt,bblly=305pt,bburx=640pt,bbury=572pt}}
      \put(270,-190){\epsfig
{file=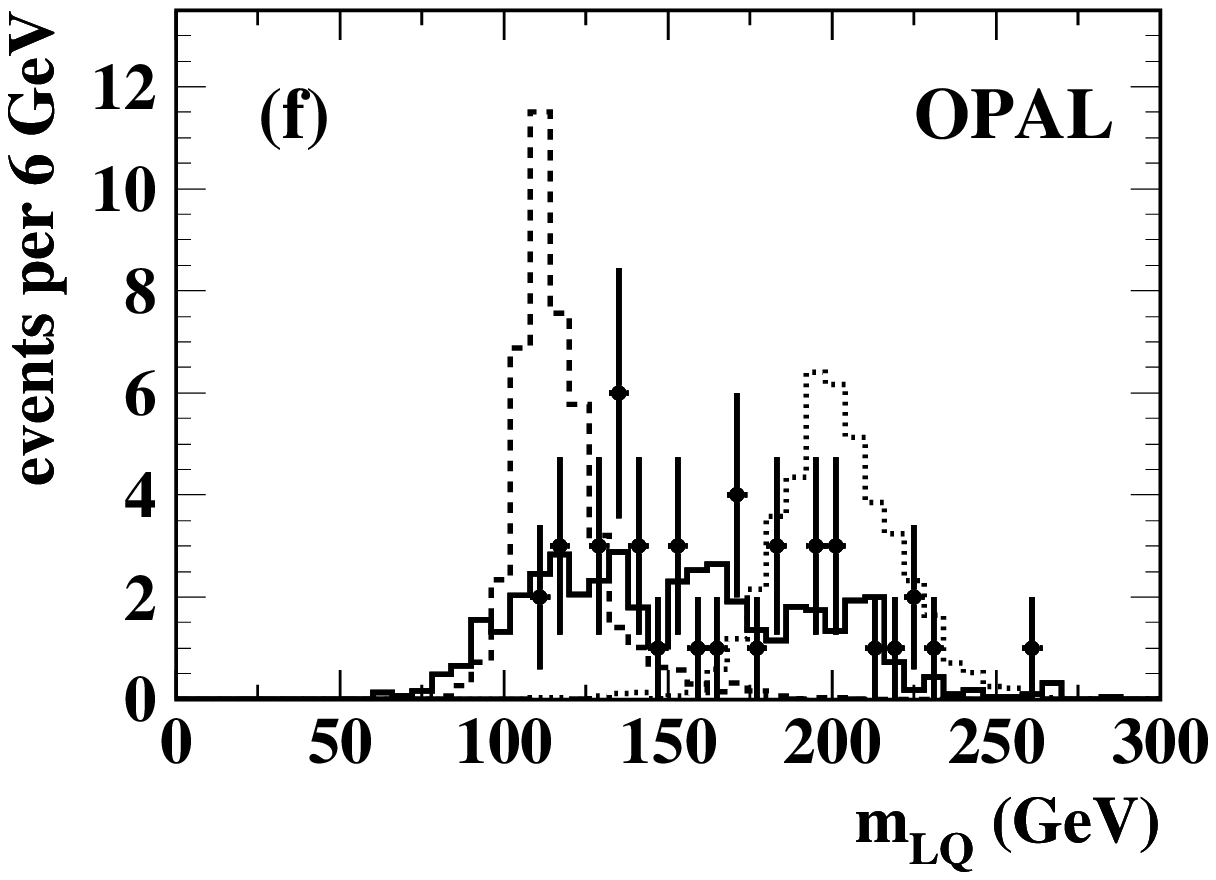,width=400pt,bbllx=20pt,bblly=305pt,bburx=640pt,bbury=572pt}}
    \end{picture}
    \vspace{125pt}
    \caption{
Electron-quark decay channel:
({\bf a}) distribution of the ratio $E_{\rm e}/\sqrt{s_{ee}}$
after cut (EQ2);
({\bf b})-({\bf d}) distributions of the likelihood input variables
$-q_{\rm e}\cos\theta_{\rm e}$, $\cos\theta^*_{\rm e}$, and $\ln\ytwothree$ after cut (EQ4);
({\bf e}) distribution of the likelihood output ${\cal L}$ after cut (EQ4);
({\bf f}) mass distribution of the selected leptoquark candidates after the full selection.
The points with error bars are the data and the full line represents the total 
Standard Model background normalised to the data luminosity.  The dashed and dotted
histograms show the distribution for the scalar state S$_{1/2}$ at $\sqrt{s_{\rm ee}}=200\,{\rm GeV}$ with scaled masses
of $x_{\rm LQ}=0.53$ and $x_{\rm LQ}=0.95$, respectively.
The normalisation of the leptoquark signals is arbitrary.
In plots (a) and (e), the arrow points into the accepted region.}
\label{eq_selection.fig}
  \end{center}
\end{figure}

Figures~\ref{eq_selection.fig}a-e show the distributions of some of 
the cut variables for data, Standard Model background and the leptoquark state 
S$_{1/2}$ with two different scaled masses $x_{\rm LQ}$.
In general, good agreement between the data and Standard Model background is observed.
The distribution of reconstructed leptoquark masses is shown in Figure~\ref{eq_selection.fig}f
for events that pass all cuts.  No significant excess is observed in the data.  

\subsection{The Neutrino Plus Hadronic Jet Channel}
\begin{figure}[htb]
  \begin{center}
    \unitlength 0.5814pt
    \begin{picture}(500,600)
      \put(-130,200){\epsfig
{file=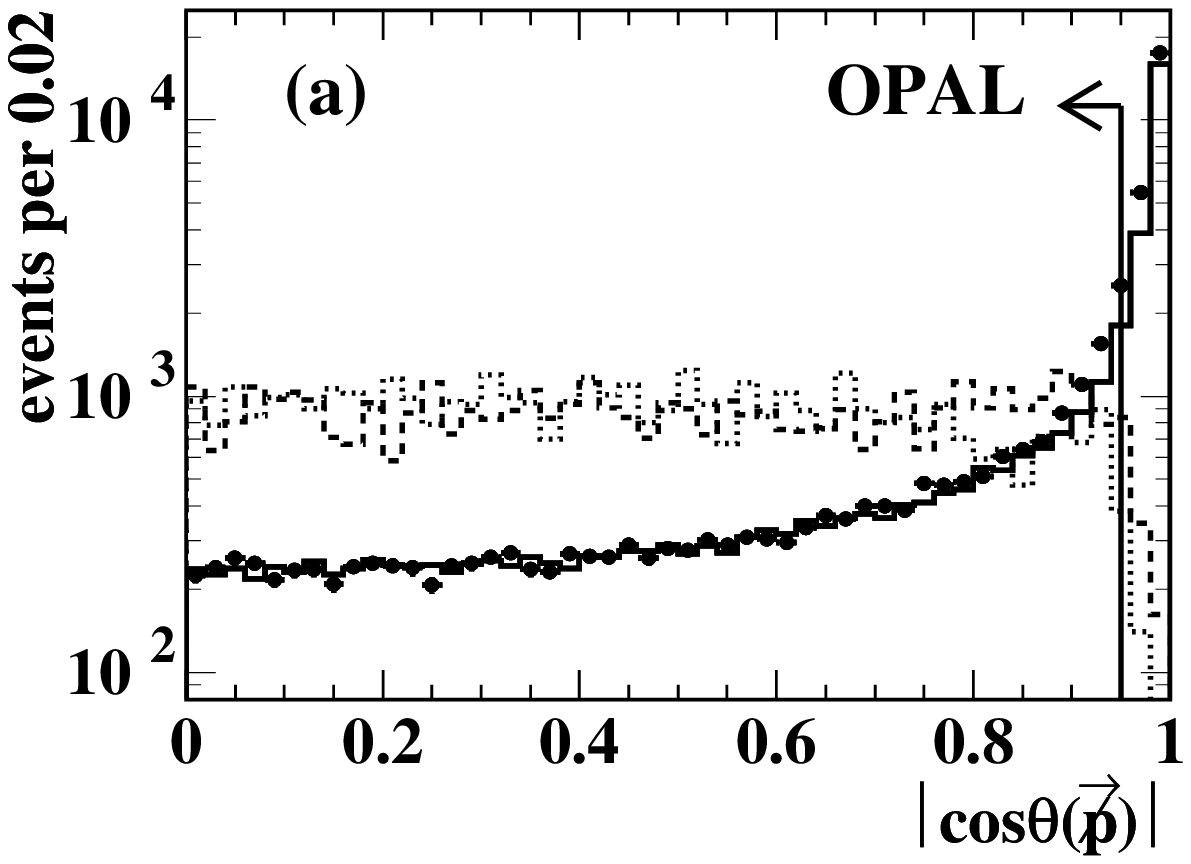,width=400pt,bbllx=20pt,bblly=305pt,bburx=640pt,bbury=572pt}}
      \put(270,200){\epsfig
{file=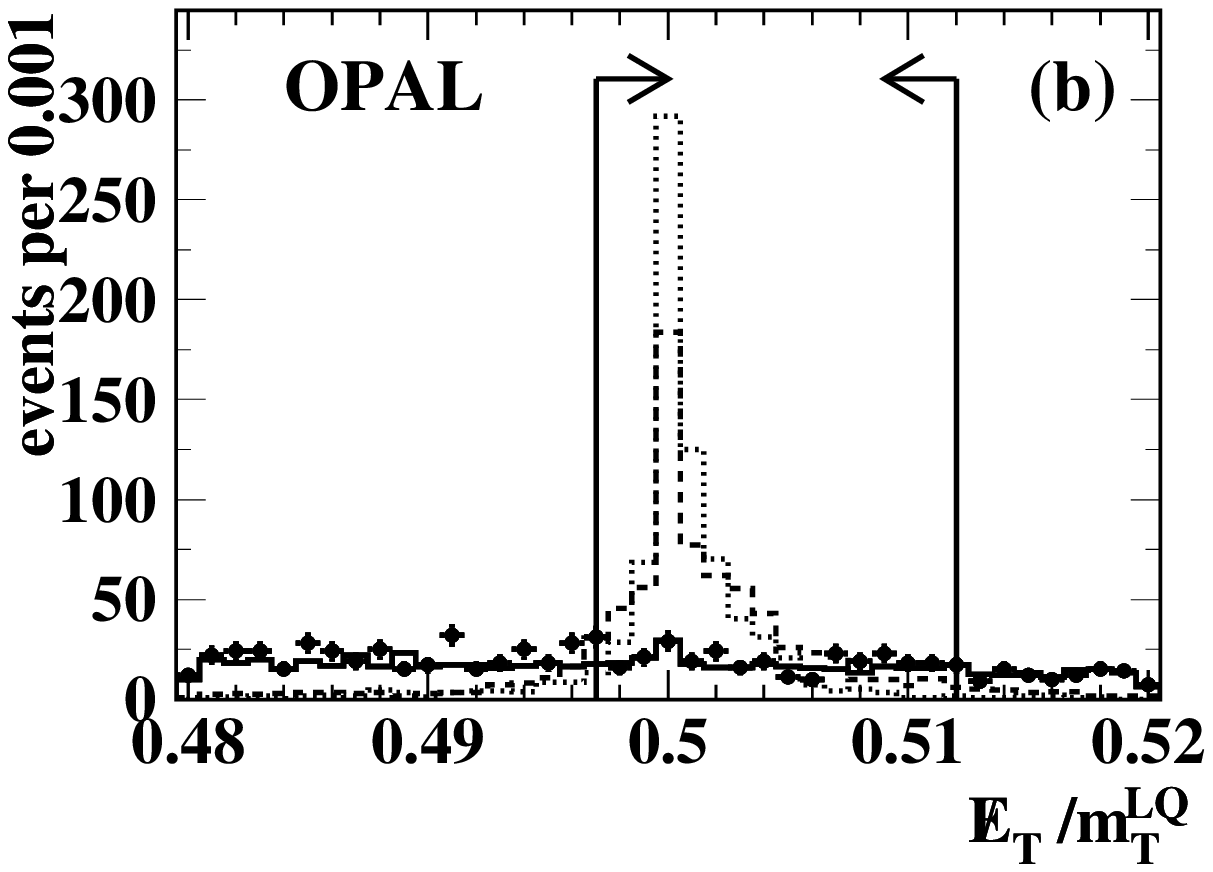,width=400pt,bbllx=20pt,bblly=305pt,bburx=640pt,bbury=572pt}}
      \put(-130,-90){\epsfig
{file=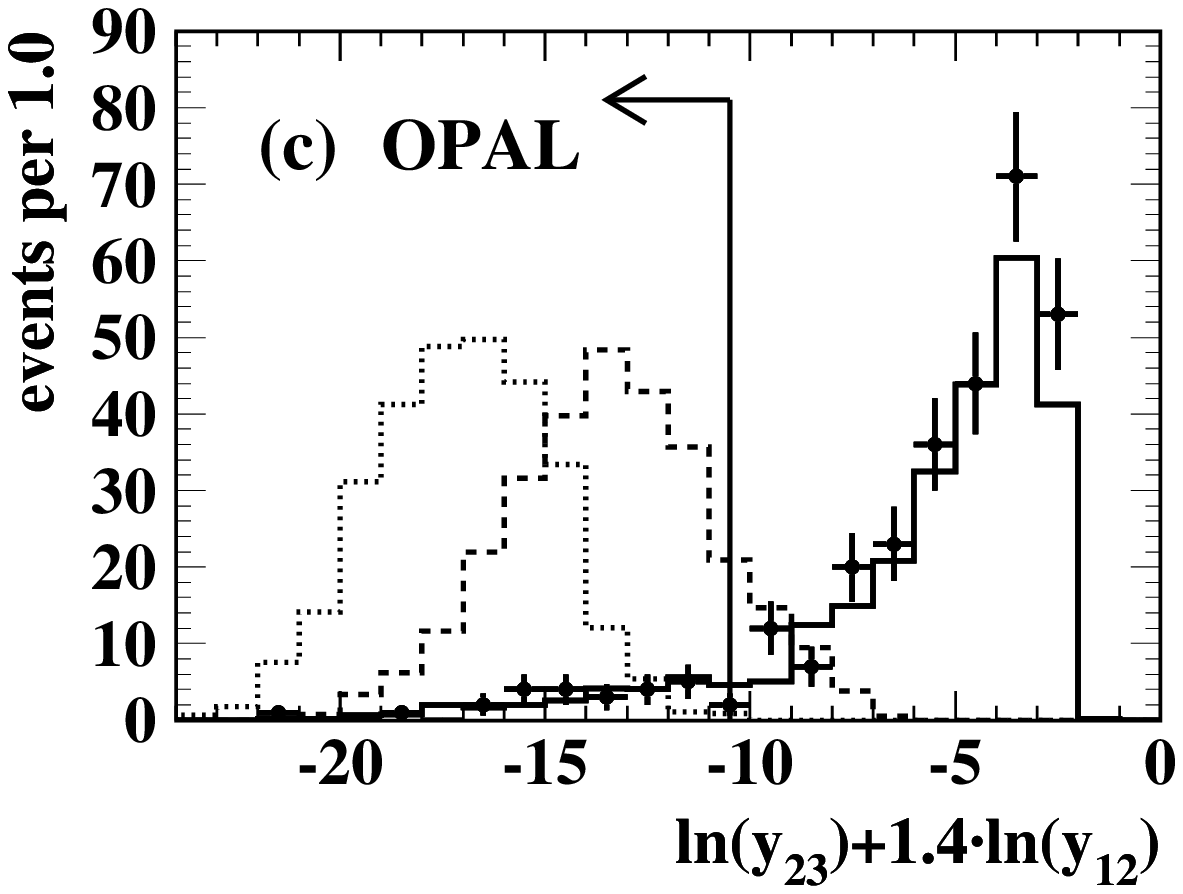,width=400pt,bbllx=20pt,bblly=305pt,bburx=640pt,bbury=572pt}}
      \put(270,-90){\epsfig
{file=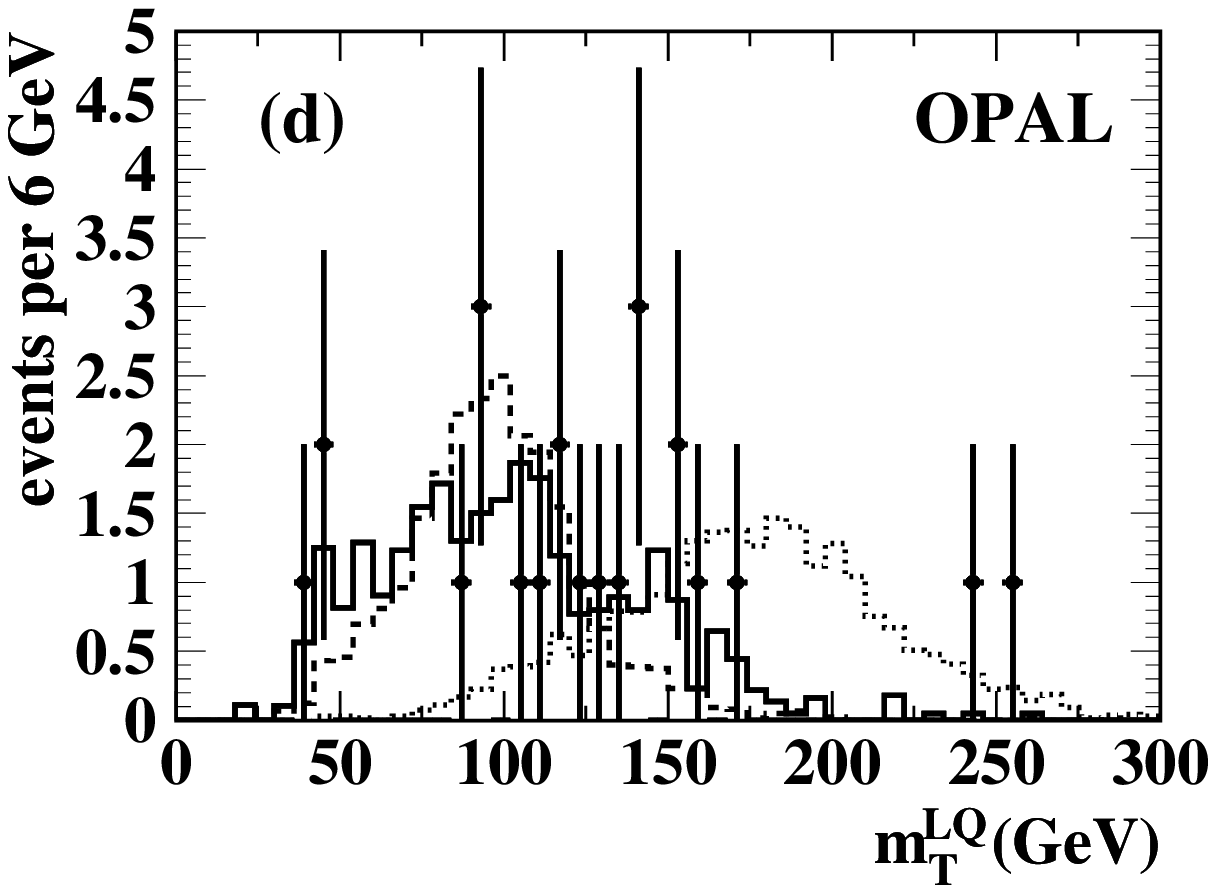,width=400pt,bbllx=20pt,bblly=305pt,bburx=640pt,bbury=572pt}}
    \end{picture}
    \vspace{70pt}
    \caption{
Neutrino-quark decay channel:
({\bf a}) 
$|\cos\theta({\vec{p}\!\!/})|$ 
distribution after cut (NQ1);
({\bf b}) distribution of the variable $\ETMISS/\mtlq$ after cut (NQ2);
({\bf c}) distribution of the variable $\ln\ytwothree+1.4\ln\yonetwo$
after cut (NQ3);
({\bf d}) transverse mass distribution of the selected leptoquark candidates after the full selection.
The points with error bars are the data and the full line represents the total 
Standard Model background normalised to the data luminosity.  The dashed and dotted
histograms show the distribution for the scalar state S$_{1/2}$ at $\sqrt{s_{\rm ee}}=200\,{\rm GeV}$ with scaled masses
of $x_{\rm LQ}=0.53$ and $x_{\rm LQ}=0.95$, respectively.
The normalisation of the leptoquark signals is arbitrary.
In plots (a), (b), and (c), the arrow points into the accepted region.}
\label{nuq_selection.fig}
  \end{center}
\end{figure}

In this case we search for events with a 
single hadronic jet whose transverse energy is balanced by missing transverse energy.
The kinematic properties of leptoquark events decaying to $\nu$q final states
do not depend strongly on the leptoquark mass, so the selection is independent of 
the reconstructed transverse mass and no likelihood is used.
The selection cuts are:
\begin{itemize}
\item[{NQ1)}]
As in the eq channel, the event must contain more than five tracks, and the energy
measured in the hadron calorimeter must exceed 1\,GeV.
\item[{NQ2)}]
The presence of at least one cone jet is required.
To ensure that the event is well contained in the detector,
the polar angle of the missing momentum vector must satisfy
$|\cos\theta({\vec{p}\!\!\!/})|<0.95$, and the polar angle of the 
most energetic cone jet
$|\cos\theta_{\rm jet\,1}|<0.85$.
\item[{NQ3)}]
The transverse mass $\mtlq$ of the leptoquark is calculated from the transverse
momentum $\pt^{\rm jet\,1}$ of the most energetic cone jet in the event
and the missing transverse momentum $\ptmiss$ of the event,
$\mtlq = \sqrt{2\pt^{\rm jet\,1}\ptmiss(1-\cos\alpha)}$, where $\alpha$ is the
opening angle between $\pt^{\rm jet\,1}$ and $\ptmiss$.
For one-jet events, there is a strong correlation between 
$\mtlq$ and $\ETMISS$, and the ratio is required to satisfy
$0.497 < \ETMISS/\mtlq < 0.512$.
The cut is chosen to be asymmetric to retain good efficiency for low-mass
leptoquarks.
\item[{NQ4)}]
Finally, the requirement of 
$\ln\ytwothree + 1.4\cdot\ln\yonetwo < -10.5$ 
is used to select events with a one-jet topology,
as shown in Figure~\ref{nuq_selection.fig}c.  
The factor of $1.4$ has been found to 
give good separation of signal and background.
\end{itemize}

\renewcommand{\arraystretch}{1.0}
\begin{table}[htbp]
  \begin{center}
    \begin{tabular}{|c||c|c|c|c|}
      \hline
               & \multicolumn{4}{|c|}{selection steps (cumulative cuts)} \\
        events & (NQ1) & (NQ2) & (NQ3) & (NQ4) \\
      \hline
      \multicolumn{5}{c}{signal efficiencies:} \\
       \hline
 scalar LQ, $m_{\rm LQ}=106.0\,{\rm GeV}$
 & $56.1\%$
 & $49.7\%$
 & $41.8\%$
 & $(35.5 \pm 4.3)\%$  
 \\
       \hline
 scalar LQ, $m_{\rm LQ}=190.5\,{\rm GeV}$
 & $66.6\%$
 & $61.7\%$
 & $56.6\%$
 & $(56.5 \pm 6.3)\%$  
 \\
       \hline
 vector LQ, $m_{\rm LQ}=106.0\,{\rm GeV}$
 & $55.1\%$
 & $48.8\%$
 & $42.5\%$
 & $(32.1 \pm 3.9)\%$  
 \\
       \hline
 vector LQ, $m_{\rm LQ}=190.5\,{\rm GeV}$
 & $59.0\%$
 & $55.9\%$
 & $51.3\%$
 & $(51.3 \pm 5.8)\%$  
 \\
       \hline
       \multicolumn{5}{c}{expected background events:} \\
       \hline
 $\epem\to{\rm 2\, fermions}$
 & $17936\phantom{.0}$ 
 & $\enspace6887\phantom{.0}$ 
 & $\enspace19.1$ 
 & $\enspace0.2\pm0.1$  
 \\
       \hline
 $\epem\to{\rm 4\, fermions}$
 & $\enspace5721\phantom{.0}$ 
 & $\enspace3886\phantom{.0}$ 
 & $226\phantom{.0}$ 
 & $20.5\pm6.6$  
 \\
       \hline
 two-photon
 & $14517\phantom{.0}$ 
 & $\enspace\enspace209\phantom{.0}$ 
 & $\enspace\enspace5.8$ 
 & $<0.4$ 
 \\
       \hline
 charged-current DIS
 & $\enspace\enspace\enspace16.9$ 
 & $\enspace\enspace\enspace12.1$ 
 & $\enspace\enspace8.8$ 
 & $\enspace6.0\pm1.9$  
 \\
       \hline
       \hline
 total expected background
 & $38192\phantom{.0}$ 
 & $10994\phantom{.0}$ 
 & $259\phantom{.0}$ 
 & $26.7\pm8.6$  
 \\
       \hline
       \multicolumn{5}{c}{observed events:} \\
       \hline
 data
 & $43094\phantom{.0}$
 & $11303\phantom{.0}$
 & $292\phantom{.0}$
 & $25$
 \\
       \hline
    \end{tabular}
    \caption{Selection efficiencies in the $\nu$q channel.
The listed efficiencies are for $\sqrt{s_{\rm ee}}=200\,{\rm GeV}$.
Also shown are the remaining number of data events and the expected
number of background events after each selection cut.
All event numbers are quoted after a loose event preselection in which 
more than 2 tracks, more than $0.1\,{\rm GeV}$ total energy in the 
hadron calorimeter, 
$|\cos\theta({\vec{p}\!\!\!/})|<0.99$, 
and 
$0.48 < \ETMISS/\mtlq < 0.52$ is required.
The errors are the total systematic uncertainties including the Monte
Carlo statistical error.  The component which 
is not due to Monte Carlo statistics is correlated between the separate 
contributions to the background.  The determination of the systematic errors 
is discussed in Section~\ref{systunc.sec}.}
    \label{cutflownuq.table}
  \end{center}
\end{table}
Table~\ref{cutflownuq.table} shows the numbers of data and expected background 
events normalised to the data luminosity after each cut.
The efficiencies for 
two different scalar and vector leptoquark masses are also given.

Figures~\ref{nuq_selection.fig}a-c show some of the selection variables for the data, 
the Standard Model 
background and the leptoquark state S$_{\rm 1/2}$ with two different
scaled masses $x_{\rm LQ}$.
The expected background describes the data well.
The transverse mass distribution of the selected leptoquark candidates is shown in
Figure~\ref{nuq_selection.fig}d after all cuts.
No significant excess is observed in the data.  Therefore, limits are calculated
on the production of leptoquarks as discussed in the following sections.

\section{Systematic Uncertainties}
\label{systunc.sec}
Systematic uncertainties arise from 
(1) the luminosity measurement, 
(2) the limited size of the simulated event samples,
(3) the modelling of the signal processes, and 
(4) the modelling of the Standard Model backgrounds.

The integrated luminosity has been varied by its uncertainty of 
$0.2\%$.  The resulting error is negligible.
The statistical error of around $1\%$ on the signal efficiencies
is taken into account.

In the simulation of signal events described in Section~\ref{sec-mc},
the leptoquark is assumed to decay before hadronisation.  However,
if the leptoquark mass and couplings to fermions are small, the
hadronisation process could begin before the leptoquark decay.  This
leads to differences in the event properties, notably in the number
of charged particle tracks per event.
Similar to~\cite{bib-karina}, 
the dependence of the efficiency on the fragmentation modelling has been studied
with a special version of {\tt PYTHIA} where the two cases have been implemented~\cite{bib-pylq}.
The resulting uncertainty in the efficiency is $5\%$
for eq~states and $11\%$ for $\nu$q~states.
Also, the parameters of the cone jet finder have been varied 
(the minimum jet energy by $(15\pm5)\,{\rm GeV}$ and the cone half angle by
($1.0\pm0.35$)
in the signal Monte Carlo.  This results in an uncertainty of $1\%$ for
eq~final states and $1\%$ - $4\%$ for $\nu$q~states.

To assess the uncertainties in the modelling of the background, 
the simulated distribution of each quantity 
which is used in the selection 
(except for the variable 
$\ln\ytwothree + 1.4\cdot\ln\yonetwo$ in the $\nu$q channel)
is reweighted in turn to match the data distribution after all preceding
cuts in the selection.  The selection (cuts and likelihood) is left unchanged.
The difference between the efficiencies for weighted and unweighted events is 
taken to be the systematic error.
       In principle, the reweighting procedure could give a 
       larger systematic error if a signal is present in the data.
       This has been checked by adding signal Monte Carlo,
       normalized to a cross-section corresponding to the
       limit of our sensitivity, to the background
       Monte Carlo and performing the reweighting procedure again.
       No change in the derived systematic error was observed.

To check the modelling of the variable $\ln\ytwothree + 1.4\cdot\ln\yonetwo$,
hadronic two-fermion events from the high-energy data have been used.  These
events have been divided along the plane perpendicular to the thrust axis,
and each half reconstructed separately.  The distributions for such half-events
have been compared between data and simulation, resulting in a $6\%$ systematic
error.

The systematic errors for all selection variables 
are added in quadrature.  The total systematic error on the 
background is $31\%$ for eq~final states and $8\%$ for $\nu$q~states.
The largest contributions are the modelling of the quantity
$-q_{\rm e}\cos\theta_{\rm e}$ ($16\%$) in the eq~case and 
$\yonetwo$ and $\ytwothree$ ($6\%$) in the $\nu$q~case.

The systematic errors on the efficiency 
are taken into account in the limit
using the procedure of Highland and Cousins~\cite{bib-syst}.
The background level is systematically shifted down by the total systematic
error ensuring conservative limits.

\section{Results}
\begin{figure}[htb]
  \begin{center}
    \begin{picture}(500,350)
      \put(-15,250){\epsfig
{file=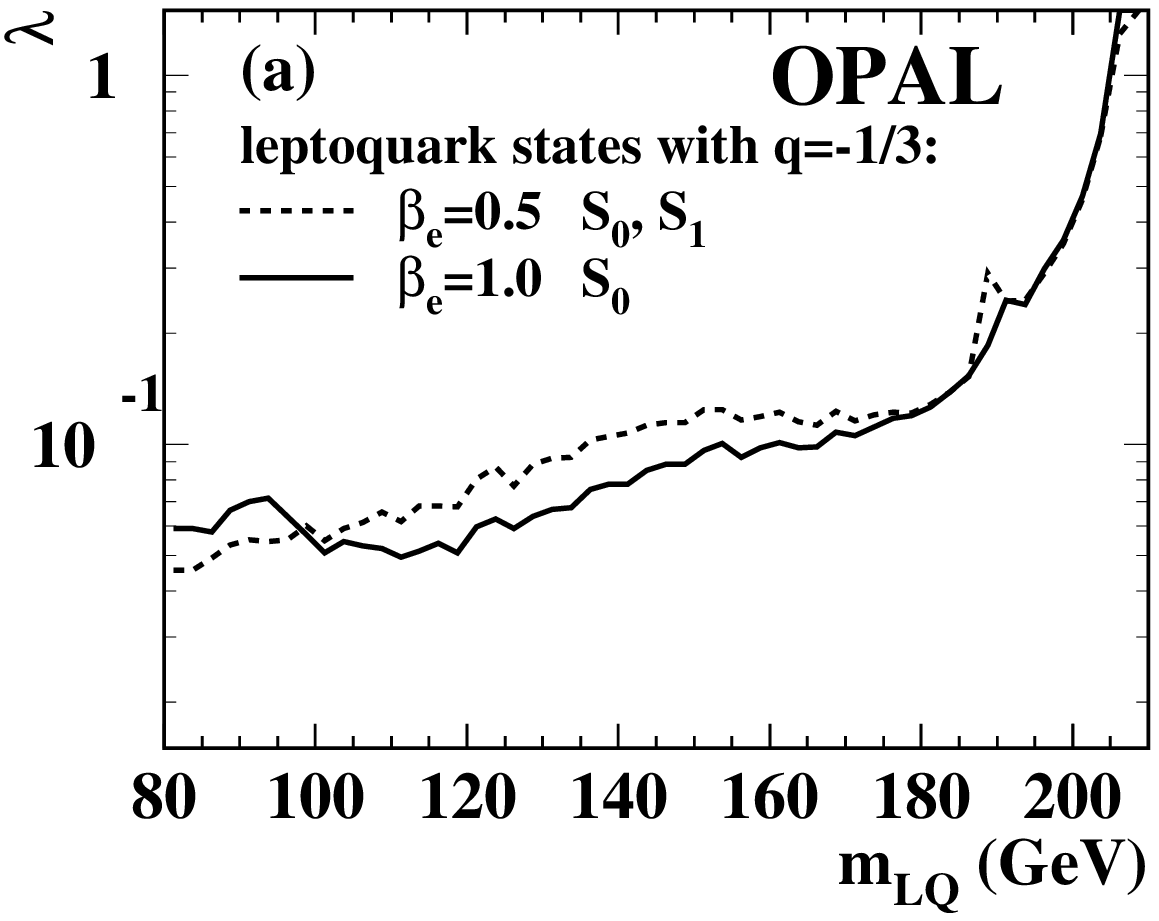,width=400pt,bbllx=20pt,bblly=305pt,bburx=640pt,bbury=572pt}}
      \put(220,250){\epsfig
{file=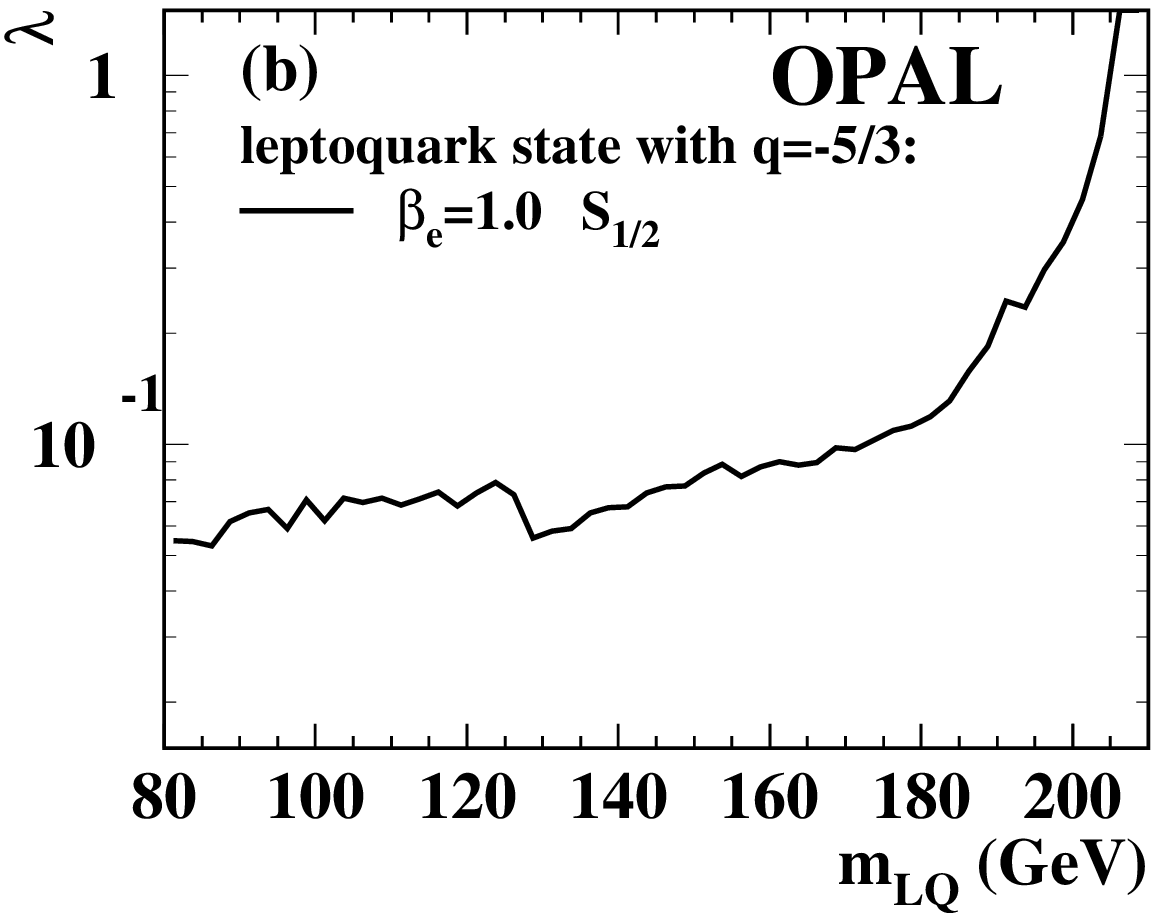,width=400pt,bbllx=20pt,bblly=305pt,bburx=640pt,bbury=572pt}}
      \put(-15,60){\epsfig
{file=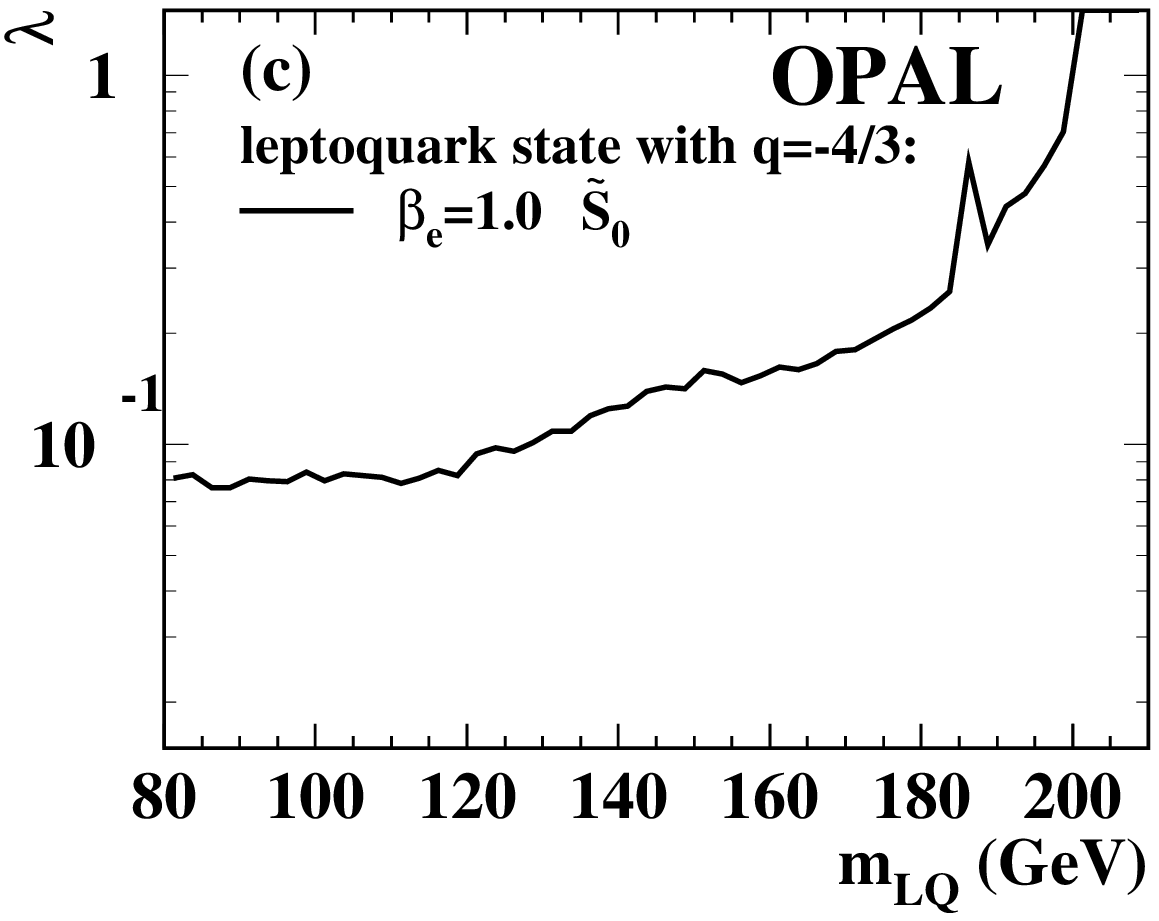,width=400pt,bbllx=20pt,bblly=305pt,bburx=640pt,bbury=572pt}}
      \put(220,60){\epsfig
{file=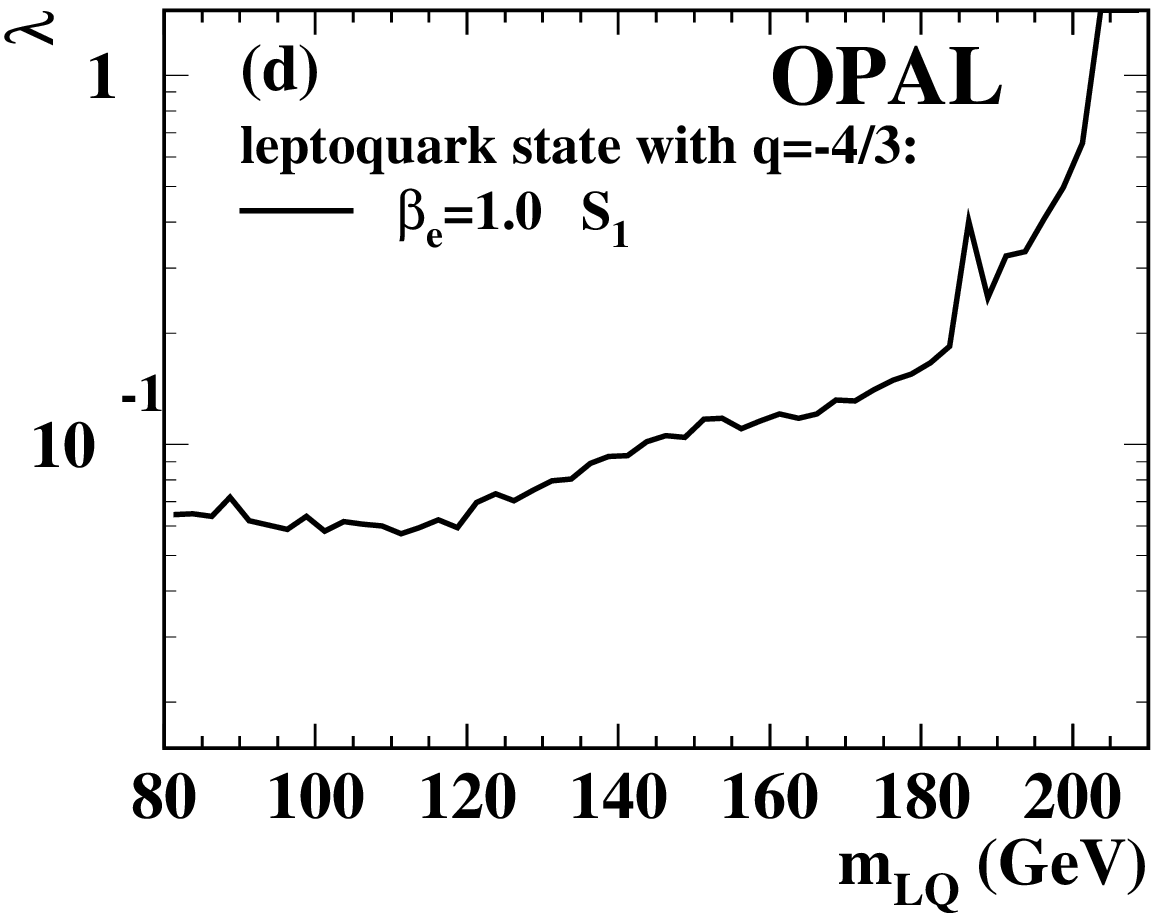,width=400pt,bbllx=20pt,bblly=305pt,bburx=640pt,bbury=572pt}}
      \put(-15,-130){\epsfig
{file=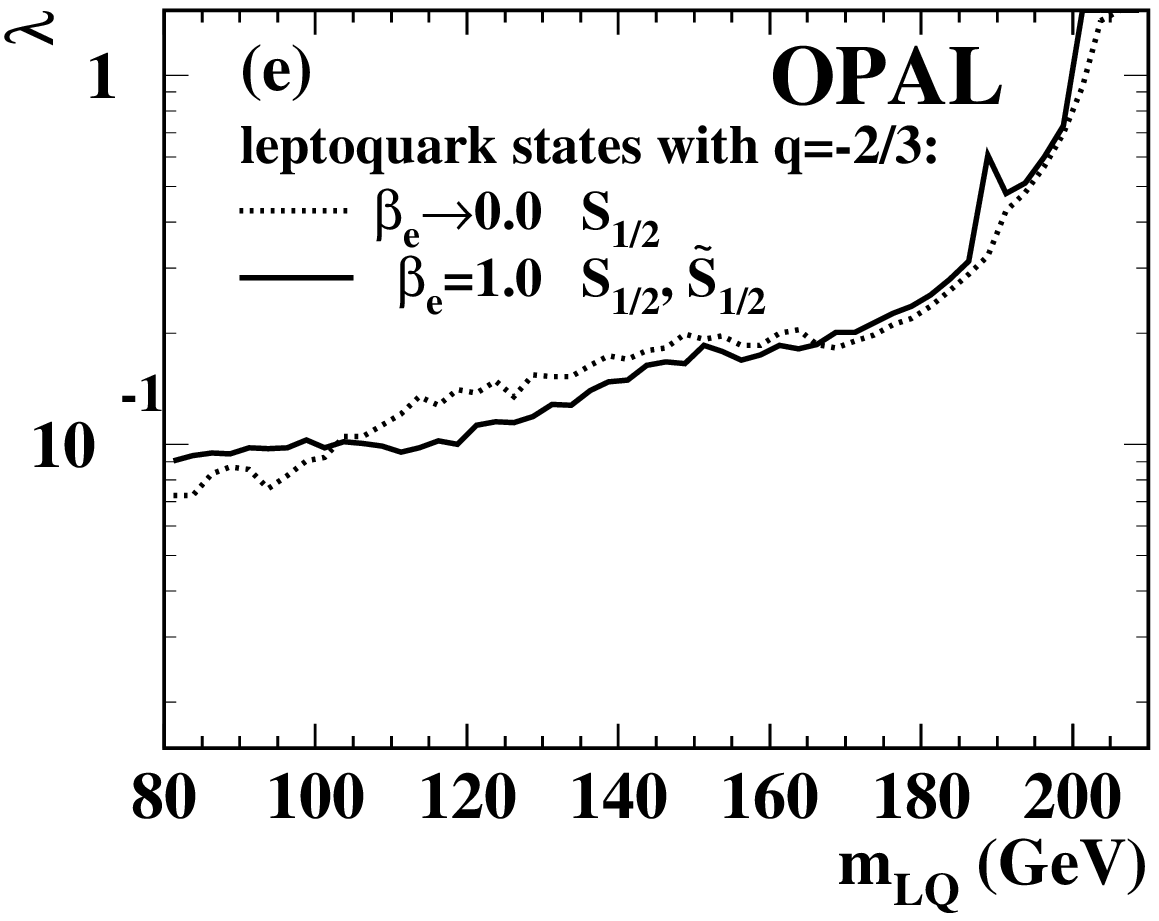,width=400pt,bbllx=20pt,bblly=305pt,bburx=640pt,bbury=572pt}}
    \end{picture}
    \vspace{150pt}
    \caption{
Upper limits at 95\% confidence level on the coupling constant $\lambda$ for single production of the scalar leptoquark states 
for different branching fractions $\brtoeq$.  
}
\label{lambda95_spin0.fig}
  \end{center}
\end{figure}

\begin{figure}[htb]
  \begin{center}
    \begin{picture}(500,350)
      \put(-15,250){\epsfig
{file=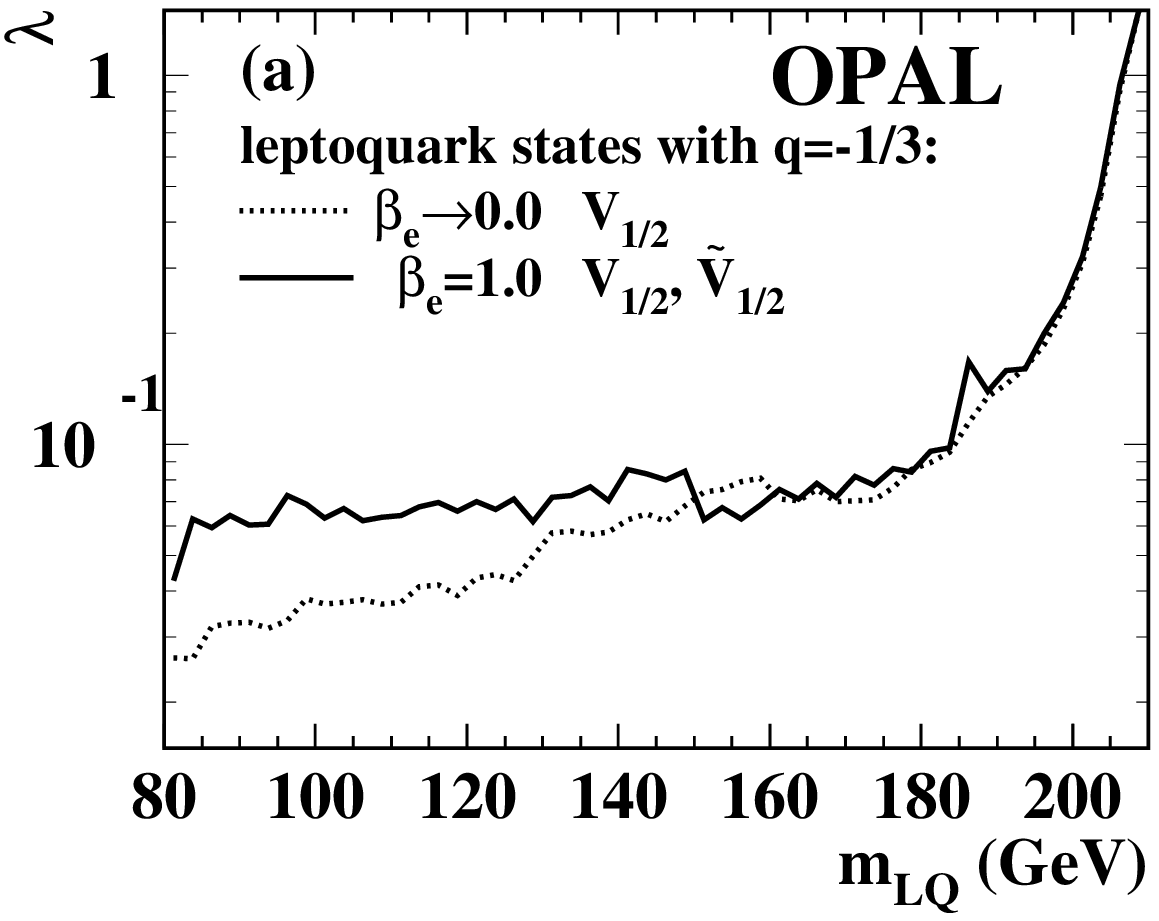,width=400pt,bbllx=20pt,bblly=305pt,bburx=640pt,bbury=572pt}}
      \put(220,250){\epsfig
{file=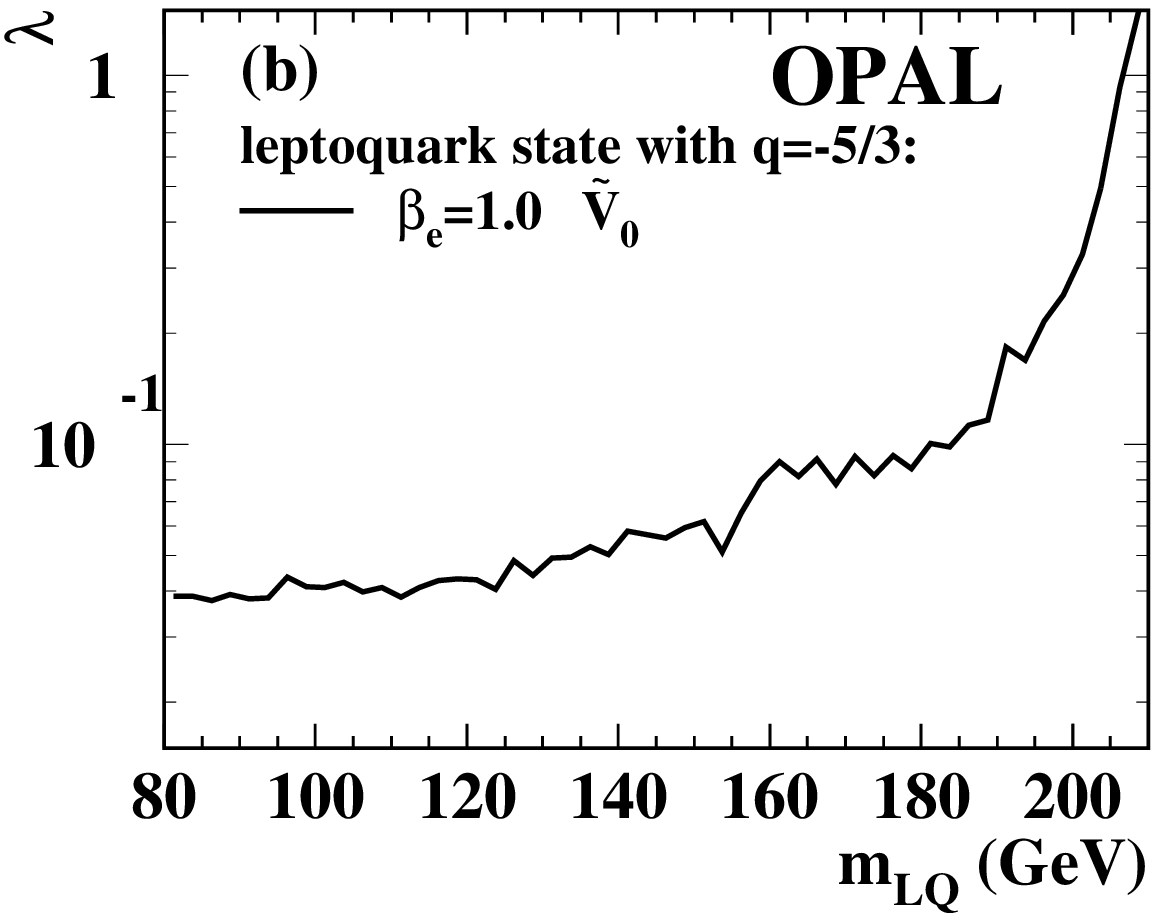,width=400pt,bbllx=20pt,bblly=305pt,bburx=640pt,bbury=572pt}}
      \put(-15,60){\epsfig
{file=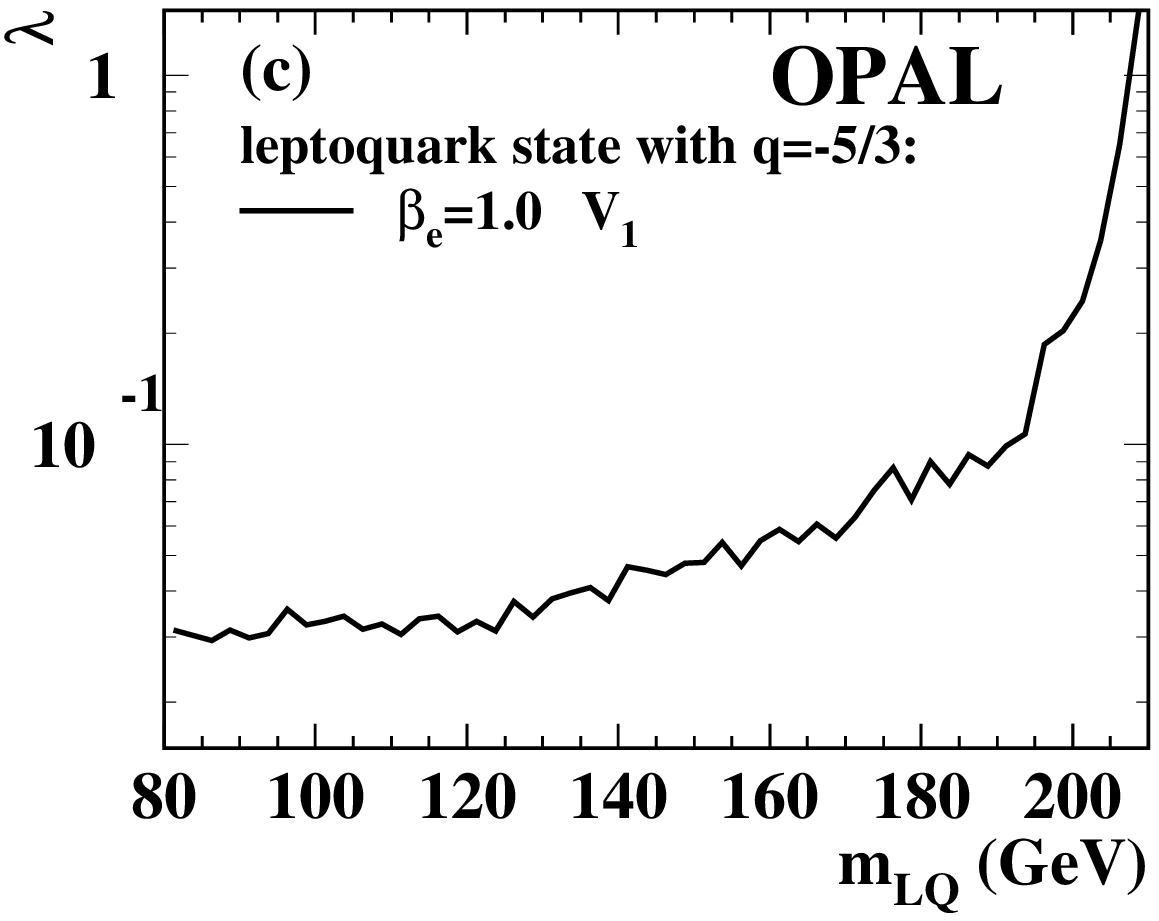,width=400pt,bbllx=20pt,bblly=305pt,bburx=640pt,bbury=572pt}}
      \put(220,60){\epsfig
{file=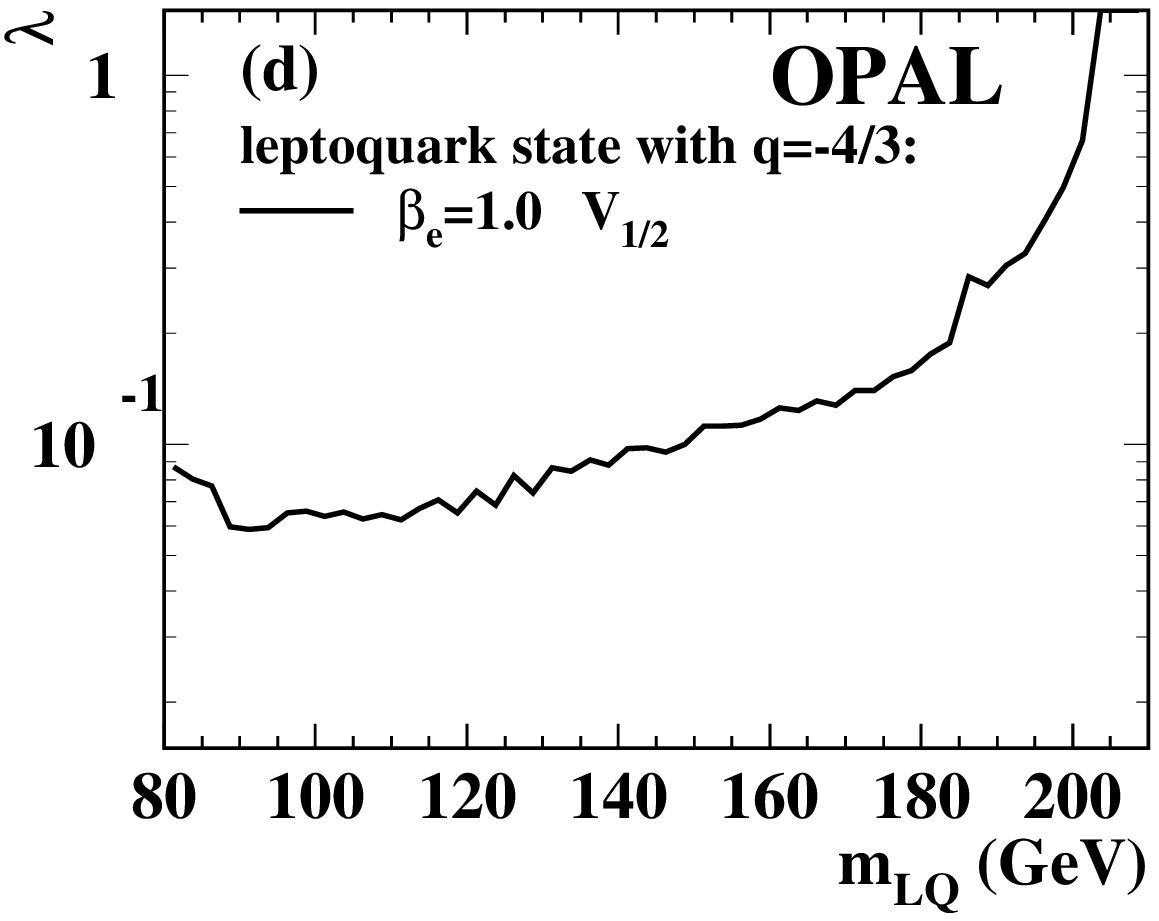,width=400pt,bbllx=20pt,bblly=305pt,bburx=640pt,bbury=572pt}}
      \put(-15,-130){\epsfig
{file=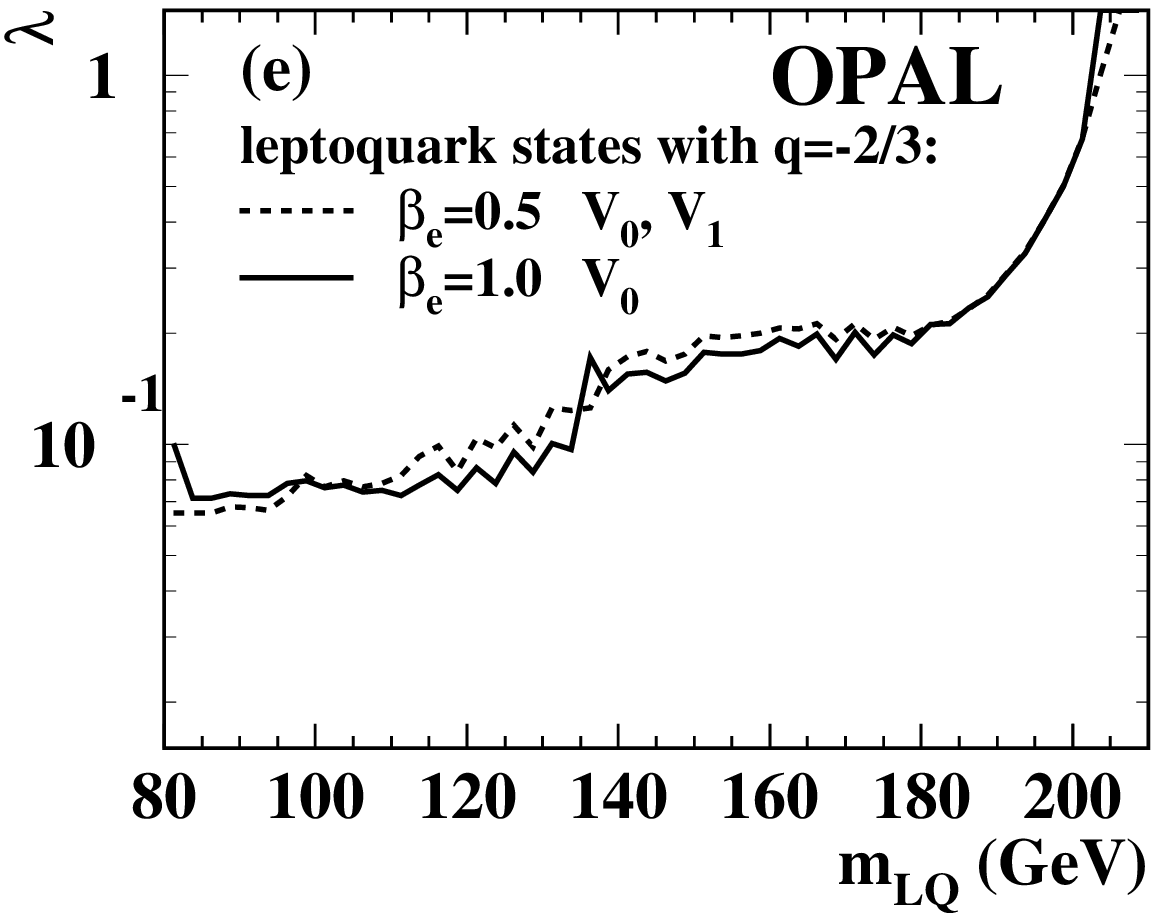,width=400pt,bbllx=20pt,bblly=305pt,bburx=640pt,bbury=572pt}}
    \end{picture}
    \vspace{150pt}
    \caption{
Upper limits at 95\% confidence level on the coupling constant $\lambda$ for single production of the vector leptoquark states 
for different branching fractions $\brtoeq$.  
}
\label{lambda95_spin1.fig}
  \end{center}
\end{figure}

Since no significant excess is observed, limits on the couplings are calculated 
for both scalar and vector leptoquarks and for those values of
the branching ratio to electron-quark final states that are allowed
in the case of chiral couplings.
The leptoquark cross-section calculated with the program {\tt ERATO-LQ} is taken
as input to obtain limits on the couplings.
The limit calculations are performed according to the
procedure of~\cite{bib-bock} which takes into account the 
expected background, the expected mass distribution, the signal efficiencies, and 
the observed candidates.
The upper limits at the $95\%$ CL of the coupling
$\lambda$ 
as a function of the mass $m_{\rm LQ}$ are
given in Figures~\ref{lambda95_spin0.fig} and~\ref{lambda95_spin1.fig}
for the scalar and vector leptoquark states 
and for different $\brtoeq$ values.  
For $\brtoeq\equiv 0$ 
no production in eq collisions is possible.  
The limits for the states S$_{\rm 1/2}$(q=$-$2/3) and V$_{\rm 1/2}$(q=$-$1/3)
for $\brtoeq\to0$ are valid if the assumption of chiral couplings is dropped.
If a coupling $\lambda=\sqrt{4\pi\alpha_{\rm em}}$ is assumed, where the electromagnetic coupling 
constant $\alpha_{\rm em}$ is taken at the
mass of the leptoquark, $\alpha_{\rm em}(M_{\rm LQ})\approx 1/128$, the mass limits
range from $183\,{\rm GeV}$ to $202\,{\rm GeV}$ depending on the leptoquark state.

In this paper, only the case of chiral couplings is considered.  If this constraint is 
dropped, then the constraints on the branching ratio $\brtoeq$ shown in Table~\ref{tab-lqstates}
no longer apply.  For a given state, the resulting mass limits do not depend strongly on $\brtoeq$.

\section{Conclusions}
\label{sec-conclusions}
We have searched for singly-produced leptoquarks in 
electron-photon interactions at e$^+$e$^-$ centre-of-mass energies
between 189 and 209\,GeV using data collected with the OPAL detector at LEP. 
No evidence is found for the production of these particles. 
Therefore, 
limits are set on the 
coupling $\lambda$ for 
scalar and vector leptoquarks
as a function of the
mass for different branching fractions $\brtoeq$ into eq final states.
The $\lambda$ limits can be directly interpreted also
as limits on 
$\lambda'$
for squarks in R-parity violating
SUSY models with the direct decay of the squark into
Standard Model particles.

\section{Acknowledgements}
We particularly wish to thank the SL Division for the efficient operation
of the LEP accelerator at all energies
 and for their close cooperation with
our experimental group.  We thank our colleagues from CEA, DAPNIA/SPP,
CE-Saclay for their efforts over the years on the time-of-flight and trigger
systems which we continue to use.  In addition to the support staff at our own
institutions we are pleased to acknowledge the  \\
Department of Energy, USA, \\
National Science Foundation, USA, \\
Particle Physics and Astronomy Research Council, UK, \\
Natural Sciences and Engineering Research Council, Canada, \\
Israel Science Foundation, administered by the Israel
Academy of Science and Humanities, \\
Minerva Gesellschaft, \\
Benoziyo Center for High Energy Physics,\\
Japanese Ministry of Education, Science and Culture (the
Monbusho) and a grant under the Monbusho International
Science Research Program,\\
Japanese Society for the Promotion of Science (JSPS),\\
German Israeli Bi-national Science Foundation (GIF), \\
Bundesministerium f\"ur Bildung und Forschung, Germany, \\
National Research Council of Canada, \\
Research Corporation, USA,\\
Hungarian Foundation for Scientific Research, OTKA T-029328, 
T023793 and OTKA F-023259,\\
Fund for Scientific Research, Flanders, F.W.O.-Vlaanderen, Belgium.\\

\end{document}